\documentclass[aps,notitlepage,twocolumn,showpacs,showkeys,preprintnumbers]{revtex4}

\usepackage{graphicx}
\usepackage{amsmath}
\usepackage{amssymb}
\usepackage{bm}
\usepackage{hyperref}

\renewcommand{\rho}{\varrho}
\renewcommand{\phi}{\varphi}
\renewcommand{\vec}{\mathbf}

\begin{document}
	
	\title{A variational atomic model of plasma accounting for ion radial correlations and electronic structure of ions (VAMPIRES)}
	
	\author{T. Blenski$^1$ and R. Piron$^{2,3}\footnote{Corresponding author, e-mail address: robin.piron@cea.fr}$}
	
	\affiliation{$^1$Laboratoire ``Interactions, Dynamiques et Lasers'', UMR 9222, CEA-CNRS-Universit\'e Paris-Saclay, Centre d'\'Etudes de Saclay, F-91191 Gif-sur-Yvette Cedex, France.}
	\affiliation{$^2$CEA, DAM, DIF, F-91297 Arpajon, France.}
	\affiliation{$^3$Universit\'{e} Paris-Saclay, CEA, Laboratoire Mati\`{e}re en Conditions Extr\^{e}mes, F-91680 Bruy\`{e}res-le-Ch\^{a}tel, France.}
	
	\date{\today}
	
	\begin{abstract}
		We propose a model of ion-electron plasma (or nucleus-electron plasma) that accounts for the electronic structure around nuclei (i.e. ion structure) as well as for ion-ion correlations. The model equations are obtained through the minimization of an approximate free-energy functional, and it is shown that the model fulfills the virial theorem. The main hypotheses of this model are 1) nuclei are treated as classical indistinguishable particles 2) electronic density is seen as a superposition of a uniform background and spherically-symmetric distributions around each nucleus (system of ions in a plasma) 3) free energy is approached using a cluster expansion (non-overlapping ions) 4) resulting ion fluid is modeled through an approximate integral equation. In the present paper, the model is described only in its average-atom version.		
	\end{abstract}
	
	
	\maketitle
	
	\section{Introduction}
The typical model of ideal plasma in thermal equilibrium is the Saha model \cite{Saha20,Saha21}. In this model, the plasma is viewed as an ideal-gas mixture, where the species are the various ion electron states, plus the free electrons. The respective shell structures of the various ion species are fixed, and calculated using a separate model of isolated ion, which can be of various degree of approximation (screened-hydrogenic, quantum detailed configuration accounting, quantum detailed level accounting...). An average-atom equivalent \cite{Mayer47} to the Saha model exists, with several possible approximations for the calculation of the average shell structure (see, for instance, \cite{Mancini85}).

In these ideal plasma models, the ion shell structure is impacted neither by the effect of the interaction between ions, nor by the effect of the polarization of continuum electrons around the ions. Such effects, whose relevance grows with density, are often designated under the generic name of ``density effects'' on the shell structure. As soon as these phenomena enter into play, the isolated-ion picture is no more valid and the issue of defining the notion of non-isolated ion, or ion in a plasma, is raised.

In order to account for these density effects, the Debye-H\"{u}ckel model can be used as a first step to improve over the ideal-gas picture. This leads to effective corrections to the isolated-ion energies: multi-electron energies in the detailed Saha model, or orbital eigenvalues in the average-atom framework (see, for instance, \cite{Griem62,Rouse62b}). Starting from the Debye-H\"{u}ckel corrections, more sophisticated corrections to the isolated-ion energies were then developed \cite{EckerWeizel56,EckerKroll63,StewartPyatt66}.

In parallel, some authors have been trying to extend the Thomas-Fermi ion-in-cell model \cite{Feynman49} to the quantum description of electrons. This led to the development of dense-plasma models such as Rozsnyai's model \cite{Rozsnyai72}, INFERNO \cite{Liberman79,Liberman82} or VAAQP \cite{Blenski07a,Blenski07b,Piron11}, in which these ``density effects'', especially pressure ionization, are built in the model to some degree of approximation.

In Rozsnyai's model, polarization of the continuum electrons is accounted for using a semi-classical (i.e. Thomas-Fermi) model for the continuum electrons, whereas bound electrons are treated through a band model, as in solid-state physics. The effects of the neighboring ions are accounted for through the neutrality of the Wigner-Seitz sphere. Due to this neutrality, the self-consistent electrostatic potential goes to zero at the Wigner-Seitz radius.

In the INFERNO model, polarization of the continuum electrons is accounted for using the same quantum formalism as for bound electrons. The effects of the neighboring ions are treated in a way similar to that of Rozsnyai's model. The Wigner-Seitz sphere is neutral, and the self-consistent electrostatic potential is zero outside the sphere.

In the VAAQP model, polarization of the continuum electrons is also accounted for using the same formalism as for bound electrons. There is no strict restriction of the potential range to the Wigner-Seitz sphere. Instead, the effects of the non-central ions are accounted for through the interaction of all electrons with a non-central-ion charge density, which is assumed to have the form of a Heaviside function. 

In all these atomic dense-plasma models, it is assumed that the non-central ions form a statistical cavity around the central ion. That is, non-central ions have zero probability to enter the Wigner-Seitz sphere. In INFERNO and VAAQP, they are uniformly distributed outside the Wigner-Seitz sphere. The relevance of this assumption comes from the qualitative behavior of the correlation function in moderately-coupled one-component classical fluids with strong repulsive interaction at short distances.

It is expected that, both in the low- and strong-coupling regimes, the ion-ion correlation function departs strongly from the cavity shape, going smoothly to the ideal-gas form in the first case, and exhibiting liquid-like structure in the second. In order to further improve these models, there is an long-lasting and ever-increasing interest for including a self-consistent description of ion-ion correlations in average-atom models \cite{Ofer88,Perrot90,Blancard04,Starrett12,Starrett13, Chihara16}. 

The potential field of application for such atomic models is wide. It includes the calculations of the equation of state and radiative properties of plasmas, the estimation of the screening impact on fusion reactions, the modeling of stopping power and X-ray Thomson scattering in dense plasmas, \textit{etc.}

When using canonical dynamics \cite{Nose84,Hoover85}, quantum-molecular-dynamics simulations may also address the ion-ion correlations in finite-temperature plasmas. Compared to such simulations, accounting for ion-ion correlations in an average-atom model is appealing for several reasons. The first is of purely scientific character: defining a notion of ion in a plasma not only allows to produce numbers, but it also endows us with an interpretation scheme of its internal functioning, when it succeeds to correctly describe the plasma.

The second reason is that, having a notion of ion in a plasma, we can extend to plasmas many of the theoretical tools of atomic physics. This notably includes the statistical approaches to the detailed accounting for excited states \cite{Perrot88,Bauche79,BarShalom89}, which are not directly applicable to molecules. This also includes the collisional-radiative modeling of plasmas out of equilibrium, which essentially resorts to the notion of atomic processes. 

Moreover, the present approach proceeds from an application of the canonical ensemble in the thermodynamic limit. On the contrary, in quantum-molecular-dynamics simulations, canonical mean values are obtained through time-averaging, while application to plasmas usually rely on periodic boundary conditions. Thermodynamic limit is then reached numerically, by increasing the size of the periodic cell.

In this paper, we propose a variational atomic model of plasma accounting for ion radial correlations and electronic structure of ions (VAMPIRES). This model defines a clear notion of ion in a plasma, and addresses the self-consistent calculation of the ion-ion correlation function and ion average electronic structure.  Unlike a previous work by the same authors \cite{Piron19b}, this model accounts for continuum electrons in the electron cloud of the ions.

The model equations are obtained through the minimization of an approximate free-energy functional, and it is shown that the model fulfills the virial theorem. This allows a rigorous approach to the thermodynamics of the system. In the present paper, the model is presented only in its average-atom version.

Unlike the broadly-used continuum-lowering models, the present model naturally introduces the screening of the self-consistent potential. It thus leads rigorously to a finite number of bound orbitals, without resorting to an \emph{ad-hoc} suppression of bound states after shifting the energies. It also accounts for the effects of the screened potential on the radial wave-functions, which is known to have an impact on the oscillator strengths (see, for instance, \cite{Shore75}) and more generally on all atomic cross-sections.

In Sec.~II, we give the general formulation of the VAMPIRES model, starting from the general many-body problem of plasma physics, and then addressing in detail each hypothesis leading to our approximate free energy. In Sec.~III, starting from the approximate free energy, we describe the variational calculation that leads to the model equations. Sec.~IV is devoted to the derivation of the various thermodynamic quantities involved in the virial theorem. A simple, analytical formula is obtained for the pressure and it is shown that the virial route to the pressure is equivalent to the usual thermodynamical one for the VAMPIRES model. In Sec. V, we comment on a first numerical application of the VAMPIRES model to the case of a Lithium plasma. Through this particular case, we stress out some peculiarities of the model. Then, we compare in Sec.~VI the results from the VAMPIRES model to those from the cavity-based models: VAAQP and INFERNO. Finally, we draw some preliminary conclusions.

	\section{General formulation of the model}
	\subsection{Classical-nuclei approximation}
	
	We consider a neutral plasma constituted of $N_\text{$\nu$}$ nuclei of charge $Z$ and $Z\,N_\text{$\nu$}$ electrons, in a
	 volume $V$. In the following, we will treat nuclei as indistinguishable classical particles, whereas electrons can be treated either quantum-mechanically or in the Thomas-Fermi approximation. In this context, we may obtain the free energy $F_\text{eq}$ of the system by minimizing a generalized-free-energy functional $F$ (see Appendix~\ref{app_notations} for the notation and Appendix~\ref{app_generalized_zwanzig} for the variational formulation of the problem):
	\begin{align}
		F_\text{eq}(N_\text{$\nu$}&,V,T)
		\nonumber\\
		=&\underset{\underline{w}}{\text{Min}}\,F\left\{\underline{w};N_\text{$\nu$},V,T\right\}
		\nonumber\\
		&\text{s.\,t. } \iint_V \frac{d^3R_1...d^3P_{N_\text{$\nu$}}}{N_\text{$\nu$}!h^{3N_\text{$\nu$}}}\left\{w(\vec{R}_1...\vec{P}_{N_\text{$\nu$}})\right\}=1
		\label{eq_Zwanzig_minimization}
	\end{align}
	where $w(\vec{R}_1...\vec{P}_{N_\text{$\nu$}})$ is a probability distribution of the nuclei classical states $(\vec{R}_1...\vec{P}_{N_\text{$\nu$}})$.
	
	The generalized-free-energy functional $F$ can be expressed as:
	\begin{widetext}
		\begin{align}
			F&\left\{\underline{w};N_\text{$\nu$},V,T\right\}\nonumber\\
			&=\iint_V \frac{d^3R_1...d^3P_{N_\text{$\nu$}}}{N_\text{$\nu$}!h^{3N_\text{$\nu$}}}\left\{
			w(\vec{R}_1...\vec{P}_{N_\text{$\nu$}})
			\left(\sum_{j=1}^{N_\text{$\nu$}} \frac{P_j^2}{2m_\text{$\nu$}}
			+F_\text{eq}^\text{e}(\vec{R}_1...\vec{R}_{N_\text{$\nu$}};N_\text{$\nu$},V,T)
			+\frac{1}{\beta}\log\left(w(\vec{R}_1...\vec{P}_{N_\text{$\nu$}})\right)
			\right)
			\right\}
		\end{align}
	\end{widetext}
	where, $F_\text{eq}^\text{e}(\vec{R}_1...\vec{R}_{N_\text{$\nu$}};N_\text{$\nu$},V,T)$ is  the free energy of electrons in the field of a fixed configuration $(\vec{R}_1...\vec{R}_{N_\text{$\nu$}})$ of the nuclei, plus the nucleus-nucleus interaction energy. 
	
	We will describe the electron free energy $F_\text{eq}^\text{e}$ using the formalism of the finite-temperature density functional theory (DFT, see \cite{Hohenberg64,Mermin65}). We thus address the calculation of $F_\text{eq}^\text{e}$ as a subsequent variational calculation:
	\begin{align}
		F_\text{eq}^\text{e}(\vec{R}_1...\vec{R}_{N_\text{$\nu$}};N_\text{$\nu$},V,T)=&\underset{\underline{n}}{\text{Min}}\,F^\text{e}\left\{\underline{n};\vec{R}_1...\vec{R}_{N_\text{$\nu$}};N_\text{$\nu$},V,T\right\}
		\nonumber\\
		&\text{ s.\,t. } \int_V d^3r\left\{n(\vec{r})\right\}=ZN_\text{$\nu$}
		\label{eq_minimiz_DFT}
	\end{align}
	where $n(\vec{r})$ is an electron density. The functional $F^\text{e}$ can be written following Kohn and Sham \cite{KohnSham65a}, as:
	\begin{align}
		F^\text{e}&\left\{\underline{n};\vec{R}_1...\vec{R}_{N_\text{$\nu$}};N_\text{$\nu$},V,T\right\}
		\nonumber\\
		=&F^0\left\{\underline{n};V,T\right\}
		+W_\text{direct}\left\{\underline{n};\vec{R}_1...\vec{R}_{N_\text{$\nu$}};N_\text{$\nu$}\right\}
		+F^\text{xc}\left\{\underline{n};V,T\right\}
	\end{align}
	Here, $F^0$ denotes the kinetic-entropic contribution to the free energy of a non-interacting electrons gas of density $n(\vec{r})$, $W_\text{direct}$ denotes the total direct-interaction energy, which includes the nucleus-nucleus contribution: 
	\begin{align}
		W_\text{direct}&\left\{\underline{n};\vec{R}_1...\vec{R}_{N_\text{$\nu$}};N_\text{$\nu$}\right\}
		\nonumber\\
		=&\frac{e^2}{2}\sum_{i=1}^N\sum_{\substack{j=1\\j\neq i}}^N\frac{Z^2}{|\vec{R}_i-\vec{R}_j|}
		+\frac{e^2}{2}\int_V d^3rd^3r'\left\{\frac{n(\vec{r})n(\vec{r}')}{|\vec{r}-\vec{r}'|}\right\}
		\nonumber\\
		&-e^2\sum_{i=1}^N\int_V d^3r\left\{\frac{Zn(\vec{r})}{|\vec{r}-\vec{R}_i|}\right\}
	\end{align}
	and $F^\text{xc}$ denotes the exchange-correlation contribution.
	The minimization problem may then be stated as:
	\begin{widetext}
		\begin{align}
			&F_\text{eq}(N_\text{$\nu$},V,T)\nonumber\\
			&=\underset{\underline{w},\underline{n}}{\text{Min}}\,
			\iint_V \frac{d^3R_1...d^3P_{N_\text{$\nu$}}}{N_\text{$\nu$}!h^{3N_\text{$\nu$}}}\left\{w
			\left(
			\sum_{j=1}^{N_\text{$\nu$}} \frac{P_j^2}{2m_\text{$\nu$}}
			+k_B T\log\left(w\right)
			+W_\text{direct}
			\left\{\underline{n};\vec{R}_1...\vec{R}_{N_\text{$\nu$}};N_\text{$\nu$}\right\}
			+F^0\left\{\underline{n};V,T\right\}
			+F^\text{xc}\left\{\underline{n};V,T\right\}
			\right)
			\right\}
			\nonumber\\
			&\hspace{0.5cm}\text{s.\,t. }\iint_V \frac{d^3R_1...d^3P_{N_\text{$\nu$}}}{N_\text{$\nu$}!h^{3N_\text{$\nu$}}}\left\{w(\vec{R}_1...\vec{P}_{N_\text{$\nu$}})\right\}=1
			\nonumber\\
			&\hspace{0.5cm}\text{s.\,t. }\int_V d^3r\left\{n(\vec{R}_1...\vec{R}_{N_\text{$\nu$}};\vec{r})\right\}=ZN_\text{$\nu$}
			\label{eq_overall_minimiz}
		\end{align}
	\end{widetext}
	where the variables over which the minimization is performed are the functions $w(\vec{R}_1...\vec{P}_{N_\text{$\nu$}})$ and $n(\vec{R}_1...\vec{R}_{N_\text{$\nu$}};\vec{r})$.

	\subsection{Notion of ion in a plasma}
We now make the most important approximation of this model. We make the \textit{Ansatz} that the equilibrium electron density $n(\vec{R}_1...\vec{R}_{N_\text{$\nu$}};\vec{r})$ belongs to the class of densities that can be written as a sum of a background contribution and identical, spherically-symmetric contributions, each corresponding to the average electron structure of an \emph{ion}:
	\begin{align}
		n(\vec{R}_1...\vec{R}_{N_\text{$\nu$}};\vec{r})
		\approx& n\left\{\underline{q},n_0;\vec{R}_1...\vec{R}_{N_\text{$\nu$}};\vec{r}\right\}\nonumber\\
		&\equiv n_0+\sum_{i=1}^N q(|\vec{r}-\vec{R}_i|)
		\label{eq_approx_average_atom}
	\end{align}
	We also postulate that $q(r)$ is a strongly-decaying function (To some extent, the latter supposition may be checked afterwards from results of the model.).
	
	In our approach, this approximation defines the notion of interacting ions in a plasma. Each ion is composed of a nucleus surrounded by a displaced-electron cloud which is spherically-symmetric on average. The ions interact with each other and with a background of ``free'' electrons that is common to all ions. This approximation notably leads to central ionic potentials, that commutes with angular momentum operators. This opens the possibility of implementing many mathematical tools of atomic physics. 
	
	In this paper we limit ourselves to identical contributions $q(r)$ of every ion. This corresponds to an average-atom description of the plasma.  
	
In the class of electron densities of Eq.~\eqref{eq_approx_average_atom}, the minimization with respect to $n(\vec{R}_1...\vec{R}_{N_\text{$\nu$}};\vec{r})$ plasma reduces to a minimization with respect to its two remaining variables $q(r)$ and $n_0$. We now rewrite each term of Eq.~\eqref{eq_overall_minimiz}, using the approximation of Eq.~\eqref{eq_approx_average_atom}.
	
	The approximation of Eq.~\eqref{eq_approx_average_atom}, leads to the following form of the neutrality condition:
	\begin{align}
		\frac{n_0}{n_\text{$\nu$}}+\int_V d^3r\left\{q(r)\right\}=Z
		\label{eq_neutrality_one-ion}
	\end{align}
	where $n_\text{$\nu$}\equiv N_\text{$\nu$}/V$.
	
	Using Eq.~\eqref{eq_approx_average_atom}, we can rewrite $W_\text{direct}$ as:
	\begin{align}
		W_\text{direct}=&\frac{Z^2e^2}{2}\sum_{i=1}^{N_\text{$\nu$}}\sum_{\substack{j=1\\j\neq i}}^{N_\text{$\nu$}} \frac{1}{|\vec{R}_i-\vec{R}_j|}
		\nonumber\\
		&-Ze^2\sum_{i=1}^{N_\text{$\nu$}}\sum_{j=1}^{N_\text{$\nu$}} \int_V d^3r \left\{ \frac{q(r)}{|\vec{r}+\vec{R}_i-\vec{R}_j|} \right\} \nonumber\\
		&+\frac{e^2}{2}\sum_{i=1}^{N_\text{$\nu$}}\sum_{j=1}^{N_\text{$\nu$}}\int_V d^3r d^3r' \left\{ \frac{q(r)q(r')}{|\vec{r}-\vec{r}'+\vec{R}_i-\vec{R}_j|} \right\} 
		\nonumber\\
		&+n_0e^2N_\text{$\nu$}\int_V d^3rd^3r'\left\{ \frac{q(r)}{|\vec{r}-\vec{r}'|}\right\} 
		\nonumber\\
		&+N_\text{$\nu$}e^2\left( \frac{n_0^2}{2n_\text{$\nu$}}-n_0Z \right)\int_V d^3r \left\{ \frac{1}{r} \right\}
	\end{align}
	Separating the diagonal terms from the off-diagonal terms of the double sums, we get:
	\begin{align}
		W_\text{direct}
		=&\frac{1}{2}\sum_{i=1}^{N_\text{$\nu$}}\sum_{\substack{j=1\\j\neq i}}^{N_\text{$\nu$}}v_\text{ii}\left\{\underline{q};|\vec{R}_i-\vec{R}_j|\right\}\nonumber\\&
		+N_\text{$\nu$}\,W_\text{intra}\left\{\underline{q};V\right\}
		+N_\text{$\nu$}\,W_\text{bg}\left\{\underline{q},n_0;V\right\}
	\end{align}
	where we have defined:
	\begin{align}
		v_\text{ii}\left\{\underline{q};R,V\right\}
		\equiv&
		\frac{Z^2e^2}{R}
		-2Ze^2\int_V d^3r \left\{ \frac{q(r)}{|\vec{r}-\vec{R}|} \right\}
		\nonumber\\&
		+e^2\int_V d^3r d^3r' \left\{ \frac{q(r)q(r')}{|\vec{r}-\vec{r}'+\vec{R}|} \right\}
		\label{eq_def_vii_direct}
	\end{align}
	\begin{align}
		W_\text{intra}\left\{\underline{q};V\right\}\equiv&
		-Ze^2\int_V d^3r \left\{ \frac{q(r)}{r} \right\}
		\nonumber\\&
		+\frac{e^2}{2}\int_V d^3r d^3r' \left\{ \frac{q(r)q(r')}{|\vec{r}-\vec{r}'|} \right\}
		\label{U_intra_r}
	\end{align}
	\begin{align}
		W_\text{bg}\left\{\underline{q},n_0;V\right\}\equiv&
		n_0e^2\int_V d^3rd^3r'\left\{ \frac{q(r)}{|\vec{r}-\vec{r}'|}\right\}
		\nonumber\\&
		+e^2\left( \frac{n_0^2}{2n_\text{$\nu$}}-n_0Z \right)\int_V d^3r \left\{ \frac{1}{r} \right\}
		\label{U_background}
	\end{align}
	$v_\text{ii}$ can be seen as an ion-ion interaction potential, $W_\text{intra}$ as an intra-ion interaction energy and $W_\text{bg}$ as the interaction energy related to the background electron density $n_0$.

	\subsection{Cluster expansion of the free energy}
For densities in the form of Eq.~\eqref{eq_approx_average_atom}, the kinetic-entropic term $F^0$ and the exchange-correlation term $F^\text{xc}$ may be approximated using a cluster expansion in the number of ions (see, for instance, \cite{Felderhof82}). Let us write the first three terms of such cluster expansion:
	\begin{align}
		F^{\bullet}&\left\{\underline{n}(\vec{r})=n_0+\sum_{i=1}^{N_\text{$\nu$}}q(|\vec{r}-\vec{R}_i|);V,T\right\}
		\nonumber\\
		=& F^{\bullet}\left\{\underline{n}(\vec{r})=n_0;V,T\right\}
		+\sum_{i=1}^{N_\text{$\nu$}} \Delta F_1^{\bullet}\left\{\underline{q},n_0,\vec{R}_i;V,T\right\}
		\nonumber\\&
		+\frac{1}{2}\sum_{i=1}^{N_\text{$\nu$}}\sum_{\substack{j=1\\j\neq i}}^{N_\text{$\nu$}} \Delta F_2^{\bullet}\left\{\underline{q},n_0,\vec{R}_i,\vec{R}_j;V,T\right\}+...
	\end{align}
	where $\bullet$ stands for either the ``$0$'' or the ``$\text{xc}$'' label, and with the definitions:
	\begin{align}
		\Delta F_1^{\bullet}&\left\{\underline{q},n_0,\vec{R};V,T\right\}
		\nonumber\\
		=&F^{\bullet}\left\{\underline{n}(\vec{r})=n_0+q(|\vec{r}-\vec{R}|);V,T\right\}\nonumber\\
		&-F^{\bullet}\left\{\underline{n}(\vec{r})=n_0;V,T\right\}
		\nonumber\\
		=&\Delta F_1^{\bullet}\left\{\underline{q},n_0;V,T\right\}
	\end{align}
	\begin{align}
		\Delta F_2^{\bullet}&\left\{\underline{q},n_0,\vec{R}_1,\vec{R}_2;V,T\right\}
		\nonumber\\
		=&F^{\bullet}\left\{\underline{n}(\vec{r})=n_0+q(|\vec{r}-\vec{R}_1|)+q(|\vec{r}-\vec{R}_2|);V,T\right\}
		\nonumber\\&
		-\Delta F_1^{\bullet}\left\{\underline{q},n_0,\vec{R}_1;V,T\right\}
		-\Delta F_1^{\bullet}\left\{\underline{q},n_0,\vec{R}_2;V,T\right\}
		\nonumber\\&
		+F^{\bullet}\left\{\underline{n}(\vec{r})=n_0;V,T\right\}
	\end{align}

	In principle, the two-ion terms could be addressed using a method such as that of \cite{GordonKim72}.
	However, it is worth noting that in the case of a local approximation, the two-ion terms are zero as soon as electronic structures are not overlapping. In the following, we will limit ourselves to the one-ion terms since ion configurations with strongly overlapping electronic structures should have rather low probabilities $w$ \cite{Mayer47}. 
	
	In the present study, the free energy contributions $\Delta F_1^{0}$ will be calculated either quantum-mechanically or using the Thomas-Fermi approximation (local density approximation to the kinetic-entropic term). The corresponding expressions are given in Appendix \ref{app_functional_derivatives}. $\Delta F_1^\text{xc}$ will be calculated using a local-density approximation (LDA) to the exchange-correlation term:
\begin{align}
\Delta F_1^\text{xc}
=\int_V d^3r \left\{f_\text{xc}(n_0+q(r),T)-f_\text{xc}(n_0,T)\right\}
\end{align}	

All the approximations above lead to the following variational formula of the free energy:
	\begin{widetext}
		\begin{align}
			F_\text{eq}(N_\text{$\nu$},V,T)=&
			\underset{\underline{q},n_0}{\text{Min}}
			\left[F^0\left\{n_0;V,T\right\}
			+F^\text{xc}\left\{n_0;V,T\right\}
			\right.\nonumber\\&\left.
			+N_\text{$\nu$}\left( 
			\Delta F_1^0\left\{\underline{q},n_0;V,T\right\}
			+\Delta F_1^\text{xc}\left\{\underline{q},n_0;V,T\right\}
			+W_\text{intra}\left\{\underline{q},V\right\}
			+W_\text{bg}(\underline{q},n_0;V)
			\right)
			\right.\nonumber\\&\left.
			+F_\text{eq}^\text{i}
			\left\{
			\underline{v}(R)=v_\text{ii}\left\{\underline{q};R,V\right\}
			;N_\text{$\nu$},V,T
			\right\}
			\right]
			\nonumber\\
			&\text{s.\,t. }\frac{n_0}{n_\text{$\nu$}}+\int_V d^3r\left\{q(r)\right\}=Z
			\label{eq_free_energy_one-ion}
		\end{align}
		where we have defined $F_\text{eq}^\text{i}\left\{\underline{v};N_\text{$\nu$},V,T\right\}$:
		\begin{align}
			F_\text{eq}^\text{i}\left\{\underline{v};N_\text{$\nu$},V,T\right\}
			=&\underset{\underline{w}}{\text{Min}}
			\int_V \frac{d^3R_1...d^3P_{N_\text{$\nu$}}}{N_\text{$\nu$}!h^{3N_\text{$\nu$}}}\left\{
			w(\vec{R}_1...\vec{P}_{N_\text{$\nu$}})
			\left(
			\sum_{j=1}^{N_\text{$\nu$}} \frac{P_j^2}{2m_\text{$\nu$}}
			+\frac{1}{2}\sum_{i=1}^{N_\text{$\nu$}}\sum_{\substack{j=1\\j\neq i}}^{N_\text{$\nu$}}v(|\vec{R}_i-\vec{R}_j|)
			+\frac{1}{\beta}\log\left(w(\vec{R}_1...\vec{P}_{N_\text{$\nu$}})\right)
			\right)
			\right\}
			\nonumber\\
			&\text{s.\,t. }\int_V \frac{d^3R_1...d^3P_{N_\text{$\nu$}}}{N_\text{$\nu$}!h^{3N_\text{$\nu$}}}\left\{w(\vec{R}_1...\vec{P}_{N_\text{$\nu$}})\right\}=1\\
			\equiv&\underset{\underline{w}}{\text{Min}}\,
			F^\text{i}\left\{\underline{w},\underline{v};N_\text{$\nu$},V,T\right\}
			\text{ s.\,t. }\int_V \frac{d^3R_1...d^3P_{N_\text{$\nu$}}}{N_\text{$\nu$}!h^{3N_\text{$\nu$}}}\left\{w(\vec{R}_1...\vec{P}_{N_\text{$\nu$}})\right\}=1
			\label{eq_general_classical_fluid}
		\end{align}
	\end{widetext}
	$F_\text{eq}^\text{i}$ is just the free energy of a homogeneous classical fluid of particles interacting through a potential $v(R)$. In Eq.~\eqref{eq_free_energy_one-ion} the interaction potential for the classical fluid is $v(R)=v_\text{ii}\left\{\underline{q};R,V\right\}$, 
	which may be considered as the effective ion-ion interaction potential stemming from our approach.
	
	\subsection{Thermodynamic limit and classical fluid of ions}
	When the neutrality condition Eq.~\eqref{eq_neutrality_one-ion} is fulfilled, the behavior of the interaction potential $v_\text{ii}\left\{\underline{q},R;V\right\}$, in the $R\rightarrow\infty$ limit is:  
	\begin{align}
		\lim_{R\rightarrow\infty}v_\text{ii}\left\{\underline{q};R,V\right\}
		&=\frac{n_0^2e^2}{n_\text{$\nu$}^2R}=\frac{Z^{*\,2}e^2}{R}
		\label{eq_lim_vii_asympt}
	\end{align}
	where we define the effective charge $Z^*\equiv n_0/n_\text{$\nu$}$.
	That is, the ions interact through a long-ranged potential having a Coulomb tail corresponding to the effective charge $Z^*$. When they are far from each other, the ions defined from the present model behave as particles of a one-component classical plasma (OCP).
	
	The interaction energy per unit volume of such a one-component system of charged particles has a logarithmic divergence in the thermodynamic limit: 
	\begin{align}
		&\frac{1}{V}\frac{1}{2}\sum_{i=1}^{N_\text{$\nu$}}\sum_{\substack{j=1\\j\neq i}}^{N_\text{$\nu$}}\frac{Z^{*\,2}e^2}{|\vec{R}_i-\vec{R}_j|}\longrightarrow \frac{n_\text{$\nu$}^2 Z^{*\,2} e^2}{2}\int_V d^3R\left\{ \frac{1}{R} \right\}
		\nonumber\\
		&\text{when $N_\text{$\nu$}\rightarrow\infty$, $V\rightarrow\infty$, with $n_\nu$ kept constant.}
	\end{align}
	However, using Eq.~\eqref{eq_neutrality_one-ion} in Eq.~\eqref{U_background}, one finds that the $W_\text{bg}$ term has the same diverging behavior, with opposite sign. When the neutrality condition Eq.~\eqref{eq_neutrality_one-ion} is fulfilled, we get:
	\begin{align}
		\frac{N_\text{$\nu$}W_\text{bg}(\underline{q},n_0,V)}{V}=&
		-\frac{n_0^2e^2}{2}\int_V d^3r\left\{ \frac{1}{r} \right\}
	\end{align}
	
	The $W_\text{bg}$ term plays the same role in the renormalization of the interaction energy, as the homogeneous neutralizing background in the OCP model. Grouping the $F_\text{eq}^\text{i}$ and $N_\text{$\nu$}W_\text{bg}$ terms, we can consider the system in the thermodynamic limit, since the divergences cancel each other. In this limit, the relevant finite quantities are the free energy per unit volume or per ion. Let us note $\bar{F}_\text{eq}^\text{i}\equiv F_\text{eq}^\text{i}/N_\text{$\nu$}$ the free energy per ion of the ion classical fluid. 
	
	Let us define the renormalized excess free energy per ion $\bar{A}_\text{eq}^\text{i}$ as follows:
	\begin{align}
		\bar{F}_\text{eq}^\text{i}+W_\text{bg}\equiv\frac{1}{\beta}\left(\log(n_\text{$\nu$}\Lambda_T^3)-1\right)+\bar{A}_\text{eq}^\text{i}
		\label{renormalization_by_U_bg}
	\end{align}
	where the first term in the right-hand-side (RHS) corresponds to the free energy of an ideal gas of density $n_\text{$\nu$}$, $\Lambda_T=h/\sqrt{2\pi m_\text{$\nu$} k_B T}$ being the classical thermal length of the nuclei. The excess term is due to interactions among ions through the potential $v_\text{ii}$ defined in Eq.~\eqref{eq_def_vii_direct}, the interaction energy being renormalized due to the presence of $W_\text{bg}$.
	
	In the thermodynamic limit, the renormalized excess free energy of a homogeneous classical fluid with arbitrary interaction potential can be exactly related to its equilibrium radial correlation function $h_\text{eq}(R)$\footnote{The radial correlation function is related to the radial pair distribution function $g(r)$ through the simple relation: $h(r)=g(r)-1$.}, through the Debye-Kirkwood charging relation \cite{Kirkwood35}\footnote{Eq.~\eqref{eq_Debye_Kirkwood_charging} corresponds to the Debye-Kirwood charging relation for a system of charged particles neutralized by a uniform background of oppositely-charged particles, yielding renormalization of the interaction energy. For particles interacting through finite-ranged potentials, there is neither a background to consider nor divergence of the interaction energy. The relation involves in this case the pair distribution function $g_\text{eq}(R)\equiv h_\text{eq}(R)+1$ instead of $h_\text{eq}(R)$.}:
	\begin{align}
		\bar{A}_\text{eq}^\text{i}=&\frac{n_\text{$\nu$}}{2}
		\int_0^\xi d\xi' \int d^3R\left\{ h^{\xi'}_\text{eq}(R) v(R) \right\}
		\label{eq_Debye_Kirkwood_charging}
	\end{align}
	where
	\begin{align}
		h_\text{eq}^\xi(|\vec{R}_2-\vec{R}_1|) \equiv& \frac{1}{n_\text{$\nu$}^2}\lim_{\substack{N\rightarrow\infty \\ V\rightarrow \infty \\ n_\text{$\nu$}\text{ cst}}} w^{\xi,(2)}_{N,V,\text{eq}}(\vec{R}_1,\vec{R}_2)-1\\
		w^{\xi,(2)}_{N,V,\text{eq}}(\vec{R}_1,\vec{R}_2)\equiv&
		\frac{1}{(N_\text{$\nu$}-2)!}
		\int\frac{d^3P_1...d^3P_{N_\text{$\nu$}}}{h^{3N_\text{$\nu$}}}
		\nonumber\\
		&\int_V d^3R_3...d^3R_{N_\text{$\nu$}} \left\{ w_\text{eq}^\xi\left(\vec{R}_1,...,\vec{P}_N \right) \right\}
	\end{align}
	Here $w_\text{eq}^\xi$ is the canonical distribution for a classical fluid in which the interaction potential is multiplied by $\xi$. $\xi$ is called a ``charging parameter'', and allows to formally ``switch on'' the interaction potential. Another well-known route to the free energy consists in integrating the system internal energy over temperature (see, for example, \cite{LandauStatisticalPhysics}, paragraph 78).

A number of approximate models of classical fluid exist, which address the calculation of the equilibrium radial correlation function through the solution of an integral equation (see \cite{HansenMcDonald}, a monograph on the subject). 
For some of these models, an analytical formula can be obtained for the free energy \cite{MoritaHiroike60,Lado73,Piron16,Blenski17,Piron19a}. Moreover, the equations of these models can be obtained from a convenient, variational formula, resorting to a generalized-free-energy functional of the radial correlation function. In such formulations, the minimization of $\bar{A}_\text{eq}^\text{i}$ with respect to the statistical distribution $w(\vec{R}_1...\vec{P}_{N_\text{$\nu$}})$ is replaced by a minimization of the approximate free-energy functional $\bar{A}_\text{eq}^\text{i\,approx}$ with respect to the radial correlation function.
	\begin{align}
		\begin{array}{l}
			\underset{\underline{w}}{\text{Min}}\,
			\bar{A}^\text{i}\left\{\underline{w},\underline{v};n_\text{$\nu$},T\right\}\\ 
			\text{s.\,t. }
			\int \frac{d^3R_1...d^3P_{N_\text{$\nu$}}}{N_\text{$\nu$}!h^{3N_\text{$\nu$}}}\left\{w\right\}=1
		\end{array}
		\longrightarrow
		\underset{\underline{h}}{\text{Min}}\,
		\bar{A}^\text{i\,approx}\left\{\underline{h},\underline{v};n_\text{$\nu$},T\right\}
	\end{align}
	Such a minimization yields both the integral equation of the approximate theory and the corresponding equilibrium value of the free energy that stems from the charging relation.
	
	In \cite{MoritaHiroike60,Lado73}, an excess-free-energy functional is derived for the hypernetted-chain (HNC) model of classical fluids. In \cite{Piron19a}, an excess-free-energy functional is obtained for the Debye-H\"{u}ckel (DH) model. The renormalized versions of these excess-free-energy functionals have, respectively, the forms:
	\begin{align}
		\bar{A}^\text{HNC}&\left\{\underline{h},\underline{v};n_\text{$\nu$},T\right\}
		\nonumber\\
		=&\frac{n_\text{$\nu$}}{2\beta}
		\int d^3R\left\{\vphantom{\frac{(R)^2}{2}}
		h(R)\beta v(R)
		\right.\nonumber\\&\left.
		+(h(R)+1)\log\left(h(R)+1\right)
		-h(R)-\frac{h(R)^2}{2}\right\}
		\nonumber\\&
		+\frac{1}{2\beta n_\text{$\nu$}}
		\int \frac{d^3k}{(2\pi)^3}\left\{n_\text{$\nu$} h_k-
		\log\left(1+n_\text{$\nu$} h_k\right)
		\right\}
		\label{eq_hnc_free_energy_renorm}
	\end{align}
	\begin{align}
		\bar{A}^{\text{DH}}&\left\{\underline{h},\underline{v};n_\text{$\nu$},T\right\}
		\nonumber\\
		=&
		\frac{n_\text{$\nu$}}{2\beta}\int d^3R \left\{h(R)\beta v(R) \right\}
		\nonumber\\&
		+\frac{1}{2\beta n_\text{$\nu$}}\int \frac{d^3k}{(2\pi)^3}
		\left\{ n_\text{$\nu$} h_{k}-\log(1+n_\text{$\nu$} h_{k}) \right\}
		\label{eq_dh_free_energy_renorm}
	\end{align}
	where $h_k$ is the Fourier transform of $h(R)$:
	\begin{align}
		&h_k\equiv \int d^3R \left\{h(R) e^{i\vec{k}.\vec{R}}\right\}
		\label{eq_def_Fourier_transform_h}
	\end{align}
	
	Using such a formalism, we finally have to solve the following minimization problem:
	\begin{align}
		\bar{F}_\text{eq}(n_\text{$\nu$},T)
		=&\underset{\underline{h},\underline{q},n_0}{\text{Min}}
		F^\text{approx}\left\{\underline{h},\underline{q},n_0;n_\text{$\nu$},T\right\}
		\nonumber\\
		&\text{s.\,t. }\frac{n_0}{n_\text{$\nu$}}+\int d^3r\left\{q(r)\right\}=Z
		\label{eq_minimization_VAMPIE_model}
	\end{align}
	with
	\begin{align}
	F^\text{approx}&\left\{\underline{h},\underline{q},n_0;n_\text{$\nu$},T\right\}
	\equiv\nonumber\\&
	\frac{1}{\beta}\left(\log(n_\text{$\nu$}\Lambda_T^3)-1\right)+\frac{f_0(n_0;T)}{n_\text{$\nu$}}
		+\frac{f_\text{xc}(n_0;T)}{n_\text{$\nu$}}
		\nonumber\\&
		+\Delta F_1^0\left\{\underline{q},n_0;T\right\}+\Delta F_1^\text{xc}\left\{\underline{q},n_0;T\right\}+W_\text{intra}\left\{\underline{q}\right\}
		\nonumber\\&
		+\bar{A}^\text{i\,approx}\left\{\underline{h},\underline{v}(R)
		=v_\text{ii}\left\{\underline{q},R\right\};n_\text{$\nu$},T\right\}
		\vphantom{\frac{1}{\beta}}
	\end{align}
	where $f_0$ denotes the free energy per unit volume of an ideal electron gas of density $n_0$ at the temperature $T$, and $f_\text{xc}$ denotes the exchange-correlation free-energy of a homogeneous electron gas of density $n_0$ at the temperature $T$.
	
	
	The minimization of Eq.~\eqref{eq_minimization_VAMPIE_model} constitutes the basis of our variational atomic model of plasma accounting for ion radial correlations and the electronic structure of ions (VAMPIRES model), in its average-atom version.

	\section{Model equations}
	In order to perform the constrained minimization of Eq.~\eqref{eq_minimization_VAMPIE_model}, we substitute for $n_0$ the following functional of $q(r)$:
	\begin{align}
		\bar{n}_0\left\{\underline{q};n_\text{$\nu$}\right\}
		\equiv n_\text{$\nu$}\left(Z-\int d^3r\left\{q(r)\right\}\right)
		\label{eq_neutrality_n0_functional}
	\end{align}
	The model equations then stem from:
	\begin{align}
		&\frac{\delta}{\delta h(r)}
		F^\text{approx}\left\{\underline{h},\underline{q},n_0=\bar{n}_0\left\{\underline{q};n_\text{$\nu$}\right\};n_\text{$\nu$},T\right\}=0 
		\label{eq_minimiz_h}		
		\\
		&\frac{\delta}{\delta q(r)}
		F^\text{approx}\left\{\underline{h},\underline{q},n_0=\bar{n}_0\left\{\underline{q};n_\text{$\nu$}\right\};n_\text{$\nu$},T\right\}=0
		\label{eq_minimiz_q}
	\end{align}
	
	The derivative with respect to the ion-ion correlation function $h(r)$ only acts on the $\bar{A}^\text{i\,approx}$ term. The minimization with respect to $h(r)$ then yields the classical-fluid integral equation corresponding to the chosen fluid model, with the interaction potential $v_\text{ii}$. If one uses the HNC free-energy functional recalled in Eq.~\eqref{eq_hnc_free_energy_renorm}, one gets the HNC closure relation, together with the Ornstein-Zernike equation which can be viewed as definition of the direct-correlation function $c(r)$:
	\begin{align}
		&c(r)=-\beta v_\text{ii}(r)-\log(h(r)+1)+h(r)
		\label{eq_fluid_integral_hnc_closure}
		\\
		&h(r)=c(r)
		+n_\text{$\nu$}\int d^3r'\left\{
		c(|\vec{r}'-\vec{r}|)h(r') 
		\right\}
		\label{eq_fluid_integral_oz}
	\end{align}
	where $v_\text{ii}(r)$ is a shorthand notation for $v_\text{ii}\left\{\underline{q},r\right\}$. If one uses the DH free-energy functional recalled in Eq.~\eqref{eq_dh_free_energy_renorm}, one gets the DH integral equation:
	\begin{align}
		h(r)=-\beta v_\text{ii}(r)
		-n_\text{$\nu$}\int d^3r'\left\{
		\beta v_\text{ii}(|\vec{r}'-\vec{r}|)h(r') 
		\right\}
		\label{eq_fluid_integral_dh}
	\end{align}
	The derivation of these equations is recalled in Appendix~\ref{app_functional_derivatives}. 
	
	The derivative with respect to the electron-cloud density $q(r)$ directly acts on $\Delta F_1^0$, $\Delta F_1^\text{xc}$, and $W_\text{intra}$ terms. It also acts on $\Delta F_1^0$, $\Delta F_1^\text{xc}$, as well as on free-energy of the homogeneous plasma $f_0$ and $f_\text{xc}$, through the dependence of $\bar{n}_0$ on $q(r)$. Moreover, the derivative with respect to $q(r)$ also acts on $\bar{A}^\text{i\,approx}$, through the dependence of $v_\text{ii}$, on $q(r)$.
	
\begin{align}
		\frac{\delta}{\delta q(r)}&
		F^\text{approx}\left\{\underline{h},\underline{q},n_0=\bar{n}_0\left\{\underline{q};n_\text{$\nu$}\right\};n_\text{$\nu$},T\right\}
		\nonumber\\
		=&\left.\frac{\delta F^\text{approx}}{\delta q(r)}\right|_{\bar{n_0}}
		+\int d^3r\left\{\left.\frac{\partial F^\text{approx}}{\partial n_0}\right|_{\bar{n_0}}\frac{\delta \bar{n_0}}{\delta q(r)}\right\}\\
		=&\left.\frac{\delta \Delta F_1^0}{\delta q(r)}\right|_{\bar{n_0}}
		+\left.\frac{\delta \Delta F_1^\text{xc}}{\delta q(r)}\right|_{\bar{n_0}}
		+\left.\frac{\delta W_\text{intra}}{\delta q(r)}\right|_{\bar{n_0}}
		\nonumber\\
		&+\int d^3r' \left\{
		\left.\frac{\delta \bar{A}^{\text{i}\,\text{approx}}}{\delta v(r')} \right|_{v_\text{ii}(r')}
		\frac{\delta v_\text{ii}(r')}{\delta q(r)}
		\right\}
				\nonumber\\
		&-n_\text{$\nu$}\left(\frac{\mu(\bar{n}_0;T)}{n_\text{$\nu$}}
		+\frac{v_\text{xc}(\bar{n}_0;T)}{n_\text{$\nu$}}
		\vphantom{\left.\frac{\partial \Delta F_1^\text{xc}}{\partial n_0}\right|_{\bar{n_0}}}\right.
		\nonumber\\
		&\left.+
		\left.\frac{\partial \Delta F_1^0}{\partial n_0}\right|_{\bar{n_0}}	
		+\left.\frac{\partial \Delta F_1^\text{xc}}{\partial n_0}\right|_{\bar{n_0}}\right)
		\label{eq_dFeqdq}
\end{align}
where we introduced the chemical potential $\mu(n;T)\equiv \partial f_0(n,T)/\partial n$ and the exchange-correlation potential $v_\text{xc}(n;T)\equiv \partial f_\text{xc}(n,T)/\partial n$.

In our approach, both in the quantum-mechanical and in the Thomas-Fermi version of the approach, the following relations hold (see Appendix \ref{app_functional_derivatives}):
\begin{align}
\frac{\delta \Delta F_1^0}{\delta q(r)}
&=\mu(n_0,T)-\bar{v}\left\{\underline{q},n_0;r;T\right\}
\label{eq_dF10_dq}\\
\frac{\partial \Delta F_1^0}{\partial n_0}
&=-\int d^3r \left\{ \bar{v}\left\{\underline{q},n_0;r;T\right\} \right\}
\label{eq_dF10dn0}
\end{align} 
where $\bar{v}\left\{\underline{q},n_0;r;T\right\}$ is the external potential leading to the electron density $n_0+q(r)$ in a non-interacting-electron system (trial potential). In this context, the difference between quantum-mechanical and Thomas-Fermi approaches is to be found in the relation between the trial potential $\bar{v}$ and the electron density.

In the quantum-mechanical approach we have:
\begin{align}
n_0+q(r) =& 2\sum_j f_\text{FD}(\varepsilon_j,\mu(n_0,T))|\psi_j(\vec{r})|^2
\nonumber\\
& +2\int \frac{d^3p}{h^3}\left\{ f_\text{FD}(\varepsilon_p,\mu(n_0,T))|\psi_\vec{p}(\vec{r})|^2 \right\}
\label{eq_dft_quantum_density}
\end{align}
\begin{align}
&\left(\frac{-\hbar^2}{2m_\text{e}}\nabla^2_\vec{r}+\bar{v}\left\{\underline{q},n_0;r;T\right\}\right)\psi_{\left|\substack{j\\\vec{p}}\right.}(\vec{r})=\varepsilon_{\left|\substack{j\\p}\right.}\psi_{\left|\substack{j\\\vec{p}}\right.}(\vec{r})
\label{eq_dft_schrodinger_1electron}
\end{align}
where $f_\text{FD}(\varepsilon,\mu)\equiv{1}/{(e^{\beta (\varepsilon-\mu)}+1)}$ is the Fermi-Dirac distribution and the factor 2 accounts for the spin degeneracy. We define $\varepsilon_p=p^2/(2m_\text{e})$. In the present case, the basis of eigen-vectors includes both a discrete part $\{\psi_j\}$  and a continuum of states $\{\psi_\vec{p}\}$ . Eq.~\eqref{eq_dft_schrodinger_1electron} can be extended straightforwardly to an electron system described using the Dirac equation instead of the Schr\"{o}dinger equation.

In the Thomas-Fermi approximation, we have:
\begin{align}
&n_0+q(r) 
\nonumber\\
&= 2\int \frac{d^3p}{h^3}\left\{f_\text{FD}
\left(
\frac{p^2}{2m_\text{e}}+\bar{v}\left\{\underline{q},n_0;r;T\right\},\mu(n_0,T)
\right)
\right\}
\label{eq_tf_density}
\end{align}

In the local density approximation (LDA) to the exchange-correlation free energy, we have:
\begin{align}
\frac{\delta \Delta F_1^\text{xc}}{\delta q(r)}
&=v_\text{xc}\left(n_0+q(r),T\right)
\label{eq_dFxc_dq}
\\
\frac{\partial \Delta F_1^\text{xc}}{\partial n_0}
&=\int d^3r \left\{
v_\text{xc}\left(n_0+q(r),T\right)
-v_\text{xc}\left(n_0,T\right)
\right\}
\label{eq_dFxc_dn0}
\end{align}

For the intra-ion direct interaction energy, we get:
\begin{align}
\frac{\delta W_\text{intra}}{\delta q(r)}
&=\frac{-Ze^2}{r}+e^2\int d^3r'\left\{\frac{q(r')}{|\vec{r}-\vec{r}'|}\right\}
\equiv v_\text{intra}\left\{\underline{q};r\right\}
\label{eq_def_v_intra}
\end{align}

As regards the classical-ion-fluid term, both for the HNC functional of Eq.~\eqref{eq_hnc_free_energy_renorm} and for the DH functional of Eq.~\eqref{eq_dh_free_energy_renorm}, we have:
\begin{align}
\frac{\delta \bar{A}^{\text{i}\,\text{approx}}\left\{\underline{h},\underline{v};n_\text{$\nu$},T\right\}}{\delta v(r)}
&=\frac{n_\text{$\nu$}}{2}h(r)\label{eq_diff_Ai_v}
\end{align}
which is to be used together with the derivative of $v_\text{ii}$:
\begin{align}
\frac{\delta v_\text{ii}(\vec{r}')}{\delta q(\vec{r})}
=2\,v_\text{intra}\left\{\underline{q};|\vec{r}'-\vec{r}|\right\}
\label{eq_dviidq}
\end{align}

Using Eqs~\eqref{eq_dF10_dq}-\eqref{eq_dviidq} in Eq.~\eqref{eq_dFeqdq}, we get:
\begin{align}
\frac{\delta}{\delta q(r)}&
F^\text{approx}\left\{\underline{h},\underline{q},n_0=\bar{n}_0\left\{\underline{q};n_\text{$\nu$}\right\};n_\text{$\nu$},T\right\}
\nonumber\\
=&
-\bar{v}\left\{\underline{q},n_0;r;T\right\}
+v_\text{el}\left\{\underline{h},\underline{q},n_0;r;T\right\}
\nonumber\\
&+v_\text{xc}\left(n_0+q(r),T\right)
-v_\text{xc}\left(n_0,T\right)
\nonumber\\
&-n_{\nu}\int d^3r' \left\{-\bar{v}\left\{\underline{q},n_0;r';T\right\}
\right.\nonumber\\
&\left.
+v_\text{xc}\left(n_0+q(r'),T\right)
-v_\text{xc}\left(n_0,T\right) \right\}
\label{eq_dFeq_dq_2}
\end{align}
where we have defined:
\begin{align}
v_\text{el}&\left\{\underline{h},\underline{q},n_0;r;T\right\}\nonumber\\
&\equiv v_\text{intra}\left\{\underline{q};r\right\}
+n_\text{$\nu$}\int d^3r' \left\{h(r')v_\text{intra}\left\{\underline{q};|\vec{r}'-\vec{r}|\right\}\right\}
\label{eq_def_v_el}
\end{align}

In view of Eq.~\eqref{eq_dFeq_dq_2}, it is useful to define $\tilde{v}_\text{el}\left\{\underline{v}_\text{el},r\right\}$ such that:	
\begin{align}
v_\text{el}(r) = \tilde{v}_\text{el}\left\{\underline{v}_\text{el},r\right\}	- n_{\nu}\int d^3r'\left\{\tilde{v}_\text{el}\left\{\underline{v}_\text{el},r'\right\}\right\}
\label{eq_wel}
\end{align}
In terms of $\tilde{v}_\text{el}$, Eq.~\eqref{eq_dFeq_dq_2} rewrites:
\begin{align}
\frac{\delta}{\delta q(r)}&
F^\text{approx}\left\{\underline{h},\underline{q},n_0=\bar{n}_0\left\{\underline{q};n_\text{$\nu$}\right\};n_\text{$\nu$},T\right\}
\nonumber\\
=&
-\bar{v}\left\{\underline{q},n_0;r;T\right\}+\tilde{v}_\text{el}(r)
\nonumber\\
&+v_\text{xc}\left(n_0+q(r),T\right)-v_\text{xc}\left(n_0,T\right)
\nonumber\\
&-n_{\nu}\int d^3r' \left\{
-\bar{v}\left\{\underline{q},n_0;r';T\right\}+\tilde{v}_\text{el}(r)
\right.\nonumber\\
&\left.
+v_\text{xc}\left(n_0+q(r'),T\right)-v_\text{xc}\left(n_0,T\right) \right\}
\label{eq_dFeq_dq_3}
\end{align}	
where $\tilde{v}_\text{el}(r)$ is a shorthand notation for $\tilde{v}_\text{el}\left\{\underline{v}_\text{el}\left\{\underline{h},\underline{q},n_0;r';T\right\},r\right\}$. 

In the RHS of Eq.~\eqref{eq_dFeq_dq_3}, the integrand is identical to the expression standing before the integral. The solution of the minimization condition Eq.~\eqref{eq_minimiz_q} is then easily found to be:	
\begin{align}
\bar{v}\left\{\underline{q},n_0;r;T\right\}=\tilde{v}_\text{el}(r)
+v_\text{xc}\left(n_0+q(r),T\right)-v_\text{xc}\left(n_0,T\right)
\label{eq_electron_scf_equation}
\end{align}

There only remains to solve Eq.~\eqref{eq_wel} for $\tilde{v}_\text{el}$, as a functional of $v_\text{el}$. It is an integral equation which can be rewritten in the Fourier space as follows:
\begin{align}
\tilde{v}_{\text{el},k} = v_{\text{el},k}
+ n_{\nu} \tilde{v}_{\text{el},k}(2\pi)^3\delta_3(\vec{k})
\label{eq_wel_deltak}
\end{align}
where $\delta_3(\vec{k})$ denotes the 3-dimensional Dirac distribution.

Let us consider the equation corresponding to Eq.~\eqref{eq_wel_deltak} in finite space. We will use a large but finite volume $V$ and the corresponding Fourier space. We define for any function $f_{V}(\vec{r})$ on $V$: 
			\begin{align}
			f_{V,\vec{k}} &= \int_{V} d^3r f_{V}(\vec{r})\exp\{i\vec{k}\vec{r}\}\nonumber\\  	
			f_{V}(\vec{r}) &= \frac{1}{V}\sum_{\vec{k}}f_{V,\vec{k}}\exp\{-i\vec{k}\vec{r}\} 
			\label{FT finite V}
			\end{align}
		The equation corresponding to Eq.~\eqref{eq_wel_deltak} is :
			\begin{align}
			(1 - n_{\nu}V\delta_{\vec{k},0})  \tilde{v}_{\text{el},V,\vec{k}} =  v_{\text{el},V,\vec{k}} 
			\label{equation for wel finite V}
		\end{align}
		where $\delta_{\vec{k},0}$ correspond to the Kroenecker symbol for a triplet of indices $\vec{k}$.
		We immediately obtain :
\begin{align}
\tilde{v}_{\text{el},V,\vec{k}} 
= \frac{v_{\text{el},V,\vec{k}}}{1 - n_{\nu}V\delta_{\vec{k},0}} 
\label{solution for wel finite V}
\end{align}
In the $V \to \infty$ limit, Eq.~\eqref{solution for wel finite V} gives 
		\begin{align}
		\tilde{v}_{\text{el},k} = 
		\begin{cases}			
			v_{\text{el},k} & \text{if } 	\vec{k}\neq 0\\
			0 					  &	\text{if }  \vec{k}= 0
		\end{cases}
		\label{alternative for wel} 		
		\end{align}	

In order to better point out the subtle difference between $v_\text{el}$ and $\tilde{v}_\text{el}$, let us now briefly study the zero-$k$ limit of $v_{\text{el},k}$. Rewriting Eqs~\eqref{U_intra_r} and \eqref{eq_def_v_el} in the Fourier space, we get:

	\begin{align}
		v_{\text{el},k}=- \frac{4\pi e^2}{k^2}
		\left(1+n_\text{$\nu$}h_k\right)
		\left(Z-q_k\right)
	\end{align}
	$q_k$ being the Fourier transform of $q(r)$.
	From the neutrality condition of Eq.~\eqref{eq_neutrality_one-ion}, we have:
	\begin{align}
		\lim_{k\rightarrow 0}\left(Z-q_k\right) = Z^*
	\end{align}
		When the minimization with respect to $h(r)$ is performed, $h(r)$ is given either by the Ornstein-Zernike relation of Eq.~\eqref{eq_fluid_integral_oz}, in the HNC case, or directly by Eq.~\eqref{eq_fluid_integral_dh}, in the DH case. 
	From the Ornstein-Zernike relation, we have :
	\begin{align}
		1+n_\text{$\nu$}h_k=\frac{1}{1-n_\text{$\nu$}c_k}
	\end{align}
	The asymptotic limit of the HNC closure Eq.~\eqref{eq_fluid_integral_hnc_closure} indeed corresponds to the DH limit, i.e.: $\lim_{k\rightarrow 0} c_k=-\beta v_{\text{ii},k}$. Both in the HNC and in the DH cases, we thus have:
	\begin{align}
		\lim_{k\rightarrow 0} \left(1+n_\text{$\nu$}h_k\right)
		=\frac{1}{1+\beta n_\text{$\nu$} v_{\text{ii},k}}
	\end{align}
	In the DH case, this equality holds for all $k$, not only in the zero-$k$ limit.
	As we have already seen in Eq.~\eqref{eq_lim_vii_asympt}, the asymptotic behavior of $v_\text{ii}(r)$ is coulombic, so that:
	\begin{align}
		\lim_{k\rightarrow 0} v_{\text{ii},k} = \frac{4\pi Z^{*\,2}e^2}{k^2}
	\end{align}
	We thus have:
	\begin{align}
		\lim_{k\rightarrow 0} \left( 1+n_\text{$\nu$}h_k \right) 
		= \frac{k^2}{k^2+4\pi n_\text{$\nu$}\beta Z^{*\,2}e^2}\equiv \frac{k^2}{k^2+k_\text{D}^2}
	\end{align}	
	where $k_\text{D}$ is the inverse Debye length. This result is related to the Stillinger-Lovett sum rules (see Chapter 10.2 of \cite{HansenMcDonald}). Consequently, we have:

\begin{align}
		\lim_{k\rightarrow 0} v_{\text{el},k}
		&=
		-\lim_{k\rightarrow 0} \frac{4\pi Z^* e^2}{k^2+4\pi n_\text{$\nu$}\beta Z^{*\,2}e^2}\\
		&=-\frac{1}{\beta n_0 }\label{eq_value_vel_k0}
\end{align}
It is worth noting that this relation is valid whatever $q(r)$, provided that $h(r)$ fulfills the fluid integral equation and $n_0$ fulfills the neutrality relation Eq.~\eqref{eq_neutrality_one-ion} (i.e. the constrained-minimization condition with respect to $h(r)$ is fulfilled).

Due to the fact that $v_{\text{el},k=0}$ is finite, for any function $g_{\vec{k}}$ which is regular at $\vec{k}=0$ we have:
	\begin{align}	
	\int \frac{d^3k}{(2\pi)^{3}} \left\{ g_{\vec{k}} \tilde{v}_{\text{el},k} \right\} &= 
	\int \frac{d^3k}{(2\pi)^{3}} \left\{ g_{\vec{k}} v_{\text{el},k} \right\}
	\\ 
	\int d^3r \left\{ g(\vec{r}) \tilde{v}_{\text{el}}(r) \right\} &= 
	\int d^3r \left\{ g(\vec{r}) v_{\text{el}}(r) \right\} 
	\label{eq_equivalence_vel_veltilde}
	\end{align}		
	since $v_{\text{el},k}$ and $\tilde{v}_{\text{el},k}$ are equal for all continuous values of $\vec{k}$ except at the point $\vec{k}=0$ which is of measure zero with respect to continuum. Eq.~\eqref{eq_equivalence_vel_veltilde} is not fulfilled, for instance, in the case of the integral:
	\begin{align}	
	\int d^3r \left\{ n_0 \tilde{v}_{\text{el}}(r) \right\} &= 0
	\neq \int d^3r \left\{ n_0 v_{\text{el}}(r) \right\}=\frac{1}{\beta}
	\end{align}
	since it involves a constant function $n_0$ which contains a Dirac distribution in the Fourier space.

	The fact that $\tilde{v}_\text{el}$	appears in the self-consistent potential, rather than $v_\text{el}$, results from the elimination of all $\vec{k}=0$ components in the interactions, due to the neutrality of our system.
	
	To summarize: minimizing $F^\text{approx}$ with respect to $h(r)$ and $q(r)$, with $n_0=\bar{n}_0\{\underline{q};n_\text{$\nu$}\}$ fulfilling the neutrality condition, we obtain the model equations that follow:
	\begin{itemize}
		\item the ion classical-fluid integral equation, that are: Eq.~\eqref{eq_fluid_integral_dh} for the DH version, Eqs~\eqref{eq_fluid_integral_oz} and ~\eqref{eq_fluid_integral_hnc_closure} for the HNC version, the interaction potential $v_\text{ii}(r)$ being defined by Eq.~\eqref{eq_def_vii_direct}.
		\item the electron self-consistent-field equation Eq.~\eqref{eq_electron_scf_equation} which is used with the definitions of $v_\text{el}(r)$ , Eq.~\eqref{eq_def_v_el}, and $\tilde{v}_\text{el}(r)$ Eq.~\eqref{alternative for wel}. The definition of the density $q(r)$ is chosen according to the chosen electron model: Eqs~\eqref{eq_dft_quantum_density} and \eqref{eq_dft_schrodinger_1electron} correspond to the quantum version, Eq.~\eqref{eq_tf_density} and to the TF version, respectively.
		\item the definition of the electron-background density $\bar{n}_0$ through the neutrality condition: Eq.~\eqref{eq_neutrality_n0_functional}.
	\end{itemize}

These are the equations of the VAMPIRES model, in its average-atom version. In the following, we will denote by the subscript ``eq'' quantities taken at the approximate equilibrium defined by $q(r)=q_\text{eq}(r;n_\text{$\nu$},T)$, $h(r)=h_\text{eq}(r;n_\text{$\nu$},T)$ fulfilling the fore-mentioned equation set. We also define $n_{0,\text{eq}}(n_\text{$\nu$},T)\equiv \bar{n}_0\left\{\underline{q}(r)=q_\text{eq}(r;n_\text{$\nu$},T);n_\text{$\nu$}\right\}$.

	\section{Thermodynamics and virial theorem}
	\subsection{Internal energy}
	The canonical internal energy per ion $\bar{U}_\text{eq}(n_\text{$\nu$},T)$ is defined as follows:
\begin{widetext}
	\begin{align}
		\bar{U}_\text{eq}\equiv&\bar{F}_\text{eq}(n_\text{$\nu$},T)-T\frac{\partial \bar{F}_\text{eq}(n_\text{$\nu$},T)}{\partial T}\\
		=&\bar{F}_\text{eq}(n_\text{$\nu$},T)
		-T
		\left(
		\left.\frac{\partial F^\text{approx}}{\partial T}\right|_\text{eq}
		+\int d^3r\left\{
		\left.\frac{\delta F^\text{approx}}{\delta h(r)}\right|_\text{eq}
		\frac{\partial h_\text{eq}(r;n_\text{$\nu$},T)}{\partial T}
		\right\}
		\right.\nonumber\\&\left.
		+\int d^3r\left\{
		\left.\left(
		\frac{\delta F^\text{approx}}{\delta q(r)}
		+\frac{\partial F^\text{approx}}{\partial n_0}
		\frac{\delta \bar{n_0}}{\delta q(r)}
		\right)\right|_\text{eq}
		\frac{\partial q_\text{eq}(r;n_\text{$\nu$},T)}{\partial T}
		\right\}	
		\right)
		\label{eq_internal_energy_interm}
		\end{align}
\end{widetext}
\begin{align}
		\bar{U}_\text{eq}=&\bar{F}_\text{eq}(n_\text{$\nu$},T)
		-T\left.\frac{\partial F^\text{approx}}{\partial T}\right|_\text{eq}
\end{align}

In Eq.~\eqref{eq_internal_energy_interm}, the integrands of the two integrals are immediately found to be zero because of the definition of equilibrium quantities $h_\text{eq}$ and $q_\text{eq}$. In the following of this section, we will directly omit terms that are zero just because of the definition of equilibrium.

	For both the HNC free-energy functional of Eq.~\eqref{eq_hnc_free_energy_renorm} and the DH free-energy functional of Eq.~\eqref{eq_dh_free_energy_renorm}, we have:
	\begin{align}
		\bar{A}^{\text{i}\,\text{approx}}-T\frac{\partial \bar{A}^{\text{i}\,\text{approx}}}{\partial T}
		=&\frac{n_\text{$\nu$}}{2}\int d^3r\left\{ h(r)v(r) \right\}\label{eq_int_Wi}\\
		\equiv& \bar{W}^\text{i\,approx}\left\{\underline{h},\underline{v};n_\text{$\nu$},T\right\}
	\end{align}
	We then have:
	\begin{align}
		\bar{U}_\text{eq}
		=&\frac{3}{2\beta}
		+\left( \frac{u_0(n_0;T)}{n_\text{$\nu$}}
		+\frac{u_\text{xc}(n_0;T)}{n_\text{$\nu$}}
		\right.\nonumber\\&
		+\Delta U_1^0\left\{\underline{q},n_0;T\right\}+\Delta U_1^\text{xc}\left\{\underline{q},n_0;T\right\}+W_\text{intra}\left\{\underline{q}\right\}
		\nonumber\\&\left.\left.
		+\bar{W}^\text{i\,approx}\left\{\underline{h},\underline{v}(r)
		=v_\text{ii}\left\{\underline{q},r\right\};n_\text{$\nu$},T\right\}
		\vphantom{\frac{u_0(n_0;T)}{n_\text{$\nu$}}}\right)\right|_\text{eq} 
		\label{eq_internal_energy}
	\end{align}
	where we have defined:
	\begin{align}
		u_0(n;T)\equiv& f_0(n;T)-T\frac{\partial f_0(n;T)}{\partial T}\\
		u_\text{xc}(n;T)\equiv& f_\text{xc}(n_0;T)-T\frac{\partial f_\text{xc}(n;T)}{\partial T}\\
		\Delta U_1^\text{xc}\left\{\underline{q},n_0;T\right\}
		\equiv&\int d^3r \left\{u_\text{xc}(n_0+q(r);T)-u_\text{xc}(n_0;T)\right\}\\
		\Delta U_1^0\left\{\underline{q},n_0;T\right\}
		\equiv& \Delta F_1^0-T\frac{\partial \Delta F_1^0}{\partial T}
	\end{align}

	\subsection{Interaction energy}
	The interaction energy $\bar{W}_\text{eq}(n_\text{$\nu$},T)$ may be defined through a differentiation of the free energy with respect to the squared electron charge (see \cite{Piron11}, and \cite{Feynman49} where this idea appeared originally). For that purpose, we treat $e^2$ as an external variable on which depend $\bar{F}_\text{eq}$, $F^\text{approx}$, $h_\text{eq}$, $q_\text{eq}$, $f_\text{xc}$, $\Delta F_1^\text{xc}$, $W_\text{intra}$, and $v_\text{ii}$.
	\begin{align}
		\bar{W}_\text{eq}
		=&e^2 \frac{\partial \bar{F}_\text{eq}(n_\text{$\nu$},T)}{\partial (e^2)}
		=e^2 \left.\frac{\partial F^\text{approx}\left\{\underline{h},\underline{q},n_0,\lambda;n_\text{$\nu$},T\right\}}{\partial (e^2)}\right|_\text{eq}
	\end{align}
	
	The classical-fluid contribution depends on $e^2$ through $v_\text{ii}$. We resort again to Eq.~\eqref{eq_diff_Ai_v}, and get:
	\begin{align}
		e^2&\frac{\partial \bar{A}^{\text{i}\,\text{approx}}\left\{\underline{h},\underline{v}(r)=v_\text{ii}(r);n_\text{$\nu$},T\right\}}{\partial (e^2)}\nonumber\\
		&=\frac{n_\text{$\nu$}}{2}\int d^3r\left\{ h(r)e^2\frac{\partial v_\text{ii}(r)}{\partial (e^2)} \right\}\\
		&=\bar{W}^\text{i\,approx}\left\{\underline{h},\underline{v}(r)=v_\text{ii}\left\{\underline{q},r\right\};n_\text{$\nu$},T\right\}
	\end{align}
	We thus obtain:
	\begin{align}
		\bar{W}_\text{eq}
		=&\frac{w_\text{xc}(n_0;T)}{n_\text{$\nu$}}
		+\Delta W_1^\text{xc}\left\{\underline{q},n_0;T\right\}+W_\text{intra}\left\{\underline{q}\right\}
		\nonumber\\&
		+\bar{W}^\text{i\,approx}\left\{\underline{h},\underline{v}(r)
		=v_\text{ii}\left\{\underline{q},r\right\};n_\text{$\nu$},T\right\}
		\label{eq_interaction_energy}
	\end{align}
	where we have defined:
	\begin{align}
		&w_\text{xc}(n;T)\equiv e^2\frac{\partial f_\text{xc}(n;T)}{\partial (e^2)}\\
		&\Delta W_1^\text{xc}\left\{\underline{q},n_0;T\right\}\nonumber\\
		&\hspace{1cm}\equiv\int d^3r \left\{w_\text{xc}(n_0+q(r);T)-w_\text{xc}(n_0;T)\right\}
	\end{align}
	
	\subsection{Pressure}
	The usual thermodynamic definition of the pressure writes:
	\begin{align}
		P_\text{thermo}(n_\text{$\nu$},T)=&n_\text{$\nu$}^2\frac{\partial \bar{F}_\text{eq}(n_\text{$\nu$},T)}{\partial n_\text{$\nu$}}
		\label{eq_def_general_thermo_pressure}
		\\
		=&n_\text{$\nu$}^2\left.\left(
		\frac{\partial F^\text{approx}}{\partial n_\text{$\nu$}}
		+\frac{\partial F^\text{approx}}{\partial n_0}
		\frac{\partial \bar{n}_0}{\partial n_\text{$\nu$}}
		\right)\right|_\text{eq}		
		\\
		=&n_\text{$\nu$}^2\left.\left(
		\frac{1}{\beta n_\text{$\nu$}}
		-\frac{f_0}{n_\text{$\nu$}^2}
		-\frac{f_\text{xc}}{n_\text{$\nu$}^2}
		+\frac{\partial \bar{A}^\text{i\,approx}}{\partial n_\text{$\nu$}}
		\right.\right.\nonumber\\&\left.\left.
		+
		\frac{\bar{n}_0}{n_\text{$\nu$}}
		\left(\frac{\mu}{n_\text{$\nu$}}
		+\frac{v_\text{xc}}{n_\text{$\nu$}}
		+\frac{\partial \Delta F_1^0}{\partial n_0}
		+\frac{\partial \Delta F_1^\text{xc}}{\partial n_0}
		\right)
		\right)\right|_\text{eq}		
	\end{align}
Using Eqs~\eqref{eq_dF10dn0}, \eqref{eq_dFxc_dn0} and \eqref{eq_electron_scf_equation}, the last two terms yield:
\begin{align}
\left.\left(\frac{\partial \Delta F_1^0}{\partial n_0}
+\frac{\partial \Delta F_1^\text{xc}}{\partial n_0}
\right)\right|_\text{eq}
=-\int d^3r \left\{ \tilde{v}_\text{el}(r) \right\}=0
\label{eq_contrib_vel_pressure}
\end{align}
Finally, we obtain the pressure formula that follows:
\begin{align}
P_\text{thermo}(n_\text{$\nu$},T)=&\frac{n_\text{$\nu$}}{\beta}
+n_\text{$\nu$}^2
\left.
\frac{\partial \bar{A}^\text{i\,approx}}{\partial n_\text{$\nu$}}
\right|_\text{eq}
\nonumber\\
&+\left(n_{0,\text{eq}}\mu(n_{0,\text{eq}},T)-f_0(n_{0,\text{eq}};T)
\right.\nonumber\\&\left.
+n_{0,\text{eq}}v_\text{xc}(n_{0,\text{eq}},T)
-f_\text{xc}(n_{0,\text{eq}};T)\right)
\label{eq_pressure_formula}
\end{align}
In Eq.~\eqref{eq_pressure_formula} RHS, the first term corresponds to the ideal-gas pressure of the ion classical fluid. The second term is the excess pressure of the ion classical fluid. Its expression depends on the chosen classical-fluid model and can be found in Appendix \ref{app_virial_classical_fluid} for the HNC and DH models. The four next terms correspond to the pressure of the homogeneous electron gas of density $n_{0,\text{eq}}$.
	
It may be worth noting that if the difference between $\tilde{v}_\text{el}$ and $v_\text{el}$ were omitted, then the integral of Eq.~\eqref{eq_contrib_vel_pressure} would yield $1/(\beta n_0)$ rather than $0$ (see Eq.~\eqref{eq_value_vel_k0}), and lead to a supplementary, unphysical, $n_\text{$\nu$}/\beta$ contribution to the pressure.
	
	An interpretation of Eq.~\eqref{eq_pressure_formula} is that the displaced electrons do not have an ideal-gas-type contribution to the pressure, since they are bound to ions. However, they contribute to the excess pressure, through the excess pressure of the ion classical fluid.
	
	\subsection{Virial theorem}
	For a non-relativistic system of particles interacting through Coulomb potential, the virial pressure may be defined as:
	\begin{align}
		P_\text{virial} \equiv \frac{2}{3}\,n_\text{$\nu$}\bar{U}_\text{eq} - \frac{1}{3}n_\text{$\nu$}\bar{W}_\text{eq}
	\end{align}
	Using this definition, the virial theorem may be stated as the equality between the virial pressure and the thermodynamic pressure of Eq.~\eqref{eq_def_general_thermo_pressure}:  $P_\text{thermo} = P_\text{virial}$. This theorem is fulfilled for the exact many-body problem, both in classical and quantum mechanics \cite{Clausius1870,Fock30}. In the case of an approximate theory, fulfilling this theorem is a property of the model, of crucial importance for providing sound thermodynamic quantities.
	
	Let us now calculate the virial pressure $P_\text{virial}$ in the case of the present model. Using respectively the expressions of Eqs~\eqref{eq_internal_energy} and \eqref{eq_interaction_energy} for the internal and interaction energies, we get:
	\begin{align}
		P_\text{virial} 
		=&\frac{n_\text{$\nu$}}{\beta}
		+\left(\frac{2}{3}u_0(n_0,T)
		+\frac{1}{3}\left(2u_\text{xc}(n_0,T)-w_\text{xc}(n_0,T)\right)
		\right.\nonumber\\&
		+\frac{2}{3}\Delta U_1^0+\frac{1}{3}\left( 2\Delta U_1^\text{xc}-\Delta W_1^\text{xc}\right)
		+\frac{1}{3}W_\text{intra}
		\nonumber\\&\left.\left.
		+\frac{1}{3}\bar{W}^\text{i\,approx}\left\{\underline{h},\underline{v}(r)=v_\text{ii}\left\{\underline{q},r\right\};n_\text{$\nu$},T\right\}\right)\right|_\text{eq}
	\end{align}
	
	In the case of a non-relativistic electron gas, we have the well-known relation:
	\begin{align}
		u_0(n;T)=\frac{3}{2}\left(n \mu(n,T)-f_0(n;T)\right)
	\end{align}
	Moreover, as regards the exchange-correlation contributions, it may be shown under broadly-valid assumptions \cite{Piron11} that:
	\begin{align}
		w_\text{xc}(n;T)=2u_\text{xc}(n;T)+3\left(f_\text{xc}(n;T)-nv_\text{xc}(n;T)\right)
	\end{align}
	We thus get:
	\begin{widetext}
		\begin{align}
			P_\text{virial} 
			=&\frac{n_\text{$\nu$}}{\beta}
			+n_0 \mu(n_0,T)-f_0(n_0;T)
			+n_0v_\text{xc}(n_0;T)-f_\text{xc}(n_0;T)
			\nonumber\\&
			+\frac{2}{3}\Delta U_1^0
			+\left( \int d^3r\left\{\left(n_0+q(r)\right)v_\text{xc}(n_0+q(r);T)-n_0v_\text{xc}(n_0;T)\right\}-\Delta F_1^\text{xc}\right)
			+\frac{1}{3}W_\text{intra}
			\nonumber\\&
			+\frac{1}{3}\bar{W}^\text{i\,approx}\left\{\underline{h},\underline{v}(R)=v_\text{ii}\left\{\underline{q},R\right\};n_\text{$\nu$},T\right\}
					\label{eq_expr_P_virial_ini}
		\end{align}
	\end{widetext}
	For both the quantum-mechanical approach and the Thomas-Fermi approximation to $\Delta F_1^0$, we can show that (see \cite{Slater33} and \cite{Piron11}, Eq.~(63)):
	\begin{align}
		\Delta U_1^0\left\{\underline{q},n_0;T\right\}
		=&\frac{1}{2}\int d^3r\left\{\left(n_0+q(r)\right)\vec{r}.\nabla_\vec{r}\bar{v}(r)\right\}
	\end{align}
	From Eq.~\eqref{eq_electron_scf_equation} we have:
	\begin{align}
		\Delta U_1^0=\frac{1}{2}\left( I_\text{el}^\text{virial}+I_\text{xc}^\text{virial}
		\right)
		\label{eq_dU10_I_virial}
	\end{align}
	where we have defined the following virial integrals:
	\begin{align}
		I_\text{el}^\text{virial}\equiv&\int d^3r\left\{\left(n_0+q(r)\right)\vec{r}.\nabla_\vec{r}\tilde{v}_\text{el}(r)\right\}
		\\
		I_\text{xc}^\text{virial}\equiv&\int d^3r\left\{\left(n_0+q(r)\right)\vec{r}.\nabla_\vec{r}v_\text{xc}(n_0+q(r))\right\}
	\end{align}
	For the exchange-correlation virial integral, integrating by part, we can show that (see Appendix \ref{app_vir_vxc}):
	\begin{align}
		I_\text{xc}^\text{virial}=-3&
		\left(
		\int d^3r\left\{(n_0+q(r))v_\text{xc}(n_0+q(r))-n_0v_\text{xc}(n_0)\right\}
		\right.\nonumber\\&\left.\vphantom{\int}
		-\Delta F_1^\text{xc}
		\label{eq_I_xc_virial}
		\right)
	\end{align}
	For the electrostatic virial integral, using the expressions of $\tilde{v}_\text{el}$ and $v_\text{ii}$ we can show that (see Appendix \ref{app_vir_vel}):
	\begin{align}
		I_\text{el}^\text{virial}
		=&-W_\text{intra}-W^\text{i\,approx}
		\nonumber\\
		&-\frac{n_\text{$\nu$}}{2}\int d^3r\left\{h(r)\vec{r}.\nabla_\vec{r}v_\text{ii}(r)\right\}
		\label{eq_I_el_virial}
	\end{align}
	
	In Eq.~\eqref{eq_I_el_virial} RHS, the third term corresponds to the virial of the ion-ion potential, relevant to the ion classical fluid. Both the HNC and the DH model fulfill the virial theorem related to the given interaction potential $v(r)$ (see Appendix \ref{app_virial_classical_fluid}). We then have:
	\begin{align}
		P^\text{i\,approx}_\text{ex}\left\{\underline{v};n_\text{$\nu$},T\right\}\equiv& n_\text{$\nu$}^2\left.\frac{\partial} {\partial n_\text{$\nu$}}
		\bar{A}^\text{i\,approx}\left\{\underline{h},\underline{v};n_\text{$\nu$},T\right\} \right|_{\text{eq}}
		\nonumber\\
		=&-\frac{n_\text{$\nu$}^2}{6}
		\int d^3r\left\{
		h_\text{eq}(r) \vec{r}.\nabla_{\vec{r}}v(r)\right\}
		\label{eq_virial_fluid}
	\end{align}
	where the $|_\text{eq}$ symbol means that $h(r)$ fulfills the integral equation corresponding to the chosen classical-fluid model, with the given interaction potential $v(r)$.
	
	Finally, using Eqs~\eqref{eq_expr_P_virial_ini}, \eqref{eq_dU10_I_virial}, \eqref{eq_I_xc_virial}, \eqref{eq_I_el_virial}, and \eqref{eq_virial_fluid}, we obtain the following expression for the virial pressure:
	\begin{align}
		P_\text{virial} 
		=&\frac{n_\text{$\nu$}}{\beta}
		+n_\text{$\nu$}^2\left.\frac{\partial} {\partial n_\text{$\nu$}}
		\bar{A}^\text{i\,approx}\left\{\underline{h},\underline{v};n_\text{$\nu$},T\right\} \right|_{\text{eq}}
		\nonumber\\&
		+n_{0,\text{eq}} \mu(n_{0,\text{eq}},T)-f_0(n_{0,\text{eq}};T)
		\nonumber\\&
		+n_{0,\text{eq}}v_\text{xc}(n_{0,\text{eq}};T)-f_\text{xc}(n_{0,\text{eq}};T)
	\end{align}
	which turns out to be identical to the expression of the thermodynamic pressure, Eq.~\eqref{eq_pressure_formula}. Thus, we just showed that the present model fulfills the virial theorem.
	
	\section{Numerical results for a Lithium plasma}

\begin{figure}[ht]
\centerline{\includegraphics[width=8cm]{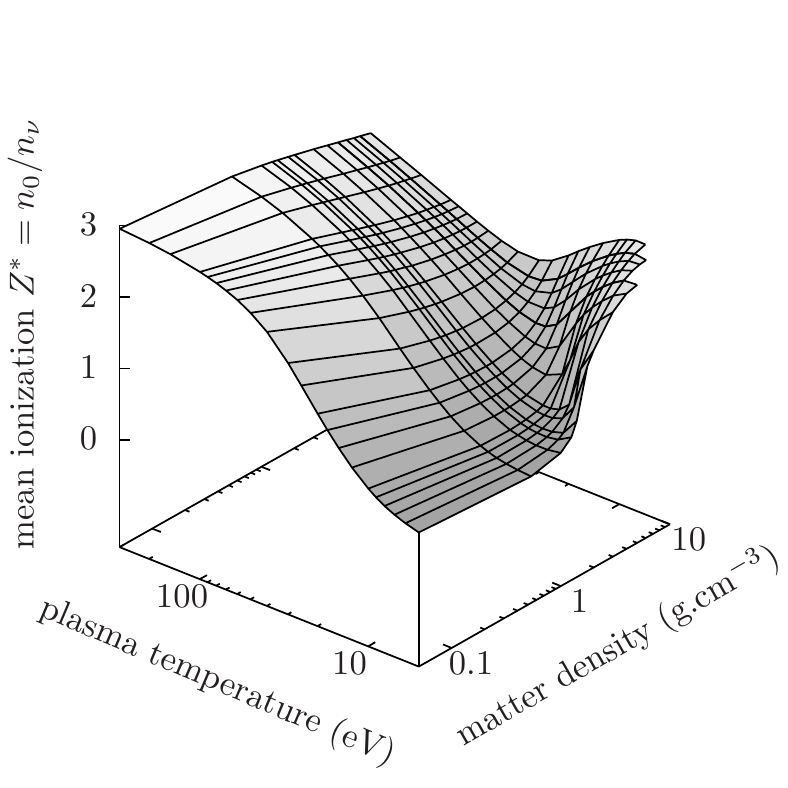}}
\caption{Lithium mean ionization from the VAMPIRES model, in its quantum average-atom version with HNC model of ion fluid and Kohn-Sham exchange-correlation term.
\label{fig_mean_ionization}}
\end{figure}

\begin{figure}[ht]
\centerline{\includegraphics[angle=90,width=8cm]{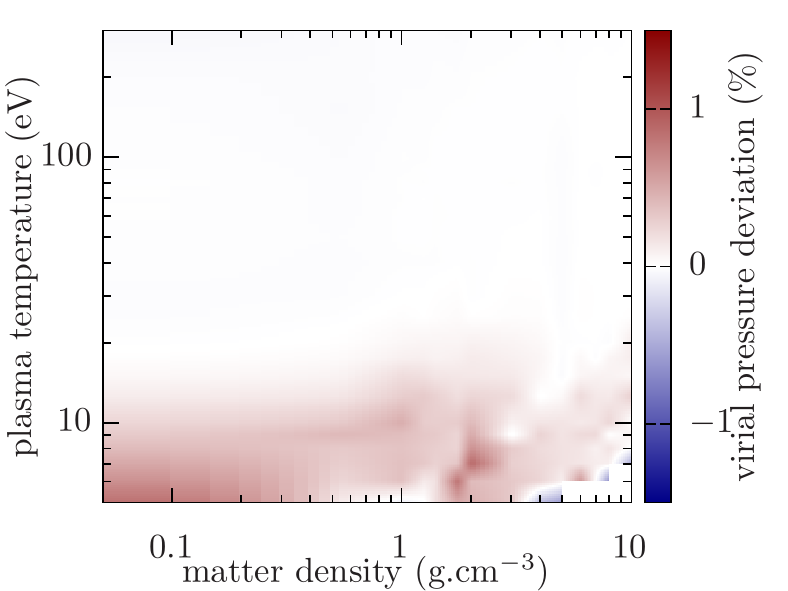}}
\caption{Map of the virial pressure deviation from the thermodynamic pressure. Values for all the results showed in this paper are between $-1\%$ and $+1\%$.
\label{fig_virial_pressure_dev}}
\end{figure}

As an illustration, we choose to apply the VAMPIRES model to the study of a Lithium plasma at densities ranging from $0.05$\,g.cm$^{-3}$ to $10$\,g.cm$^{-3}$ and temperatures ranging from 5\,eV to 300\,eV. The choice of Lithium was made for two reasons. First, as it is a light element, the ion contribution to the thermodynamic quantities are relatively high (it contributes for 1/4th in the ideal-plasma limit). Second, it turns out that numerical calculations are most difficult for strongly coupled plasma, and Lithium offers a case of moderate coupling around solid density, even at rather low temperatures.

Figure \ref{fig_mean_ionization} is a plot of the mean ionization from the VAMPIRES model, in its quantum version, with HNC model of the ion fluid, and using Kohn-Sham exchange term \cite{KohnSham65a}. Here the mean ionization $Z^*$ is defined as in Eq.~\eqref{eq_lim_vii_asympt}, from the density of the homogeneous electron background. For reasons that will be sketched below, we did not succeed in performing the calculations below 6 eV for matter densities above 5\,g.cm$^{-3}$, and below 7\,eV for matter densities above 8\,g.cm$^{-3}$.

Figure \ref{fig_virial_pressure_dev} shows the relative deviation of the virial pressure with respect to the thermodynamic pressure in our numerical calculations. As can be seen on the figure, for all the results that are presented, we have an agreement better than $1\%$ among these two values of the pressure. This gives us an idea of the numerical precision in this series of calculations. Relative precision on the virial pressure is usually worse at low temperature, because pressure values are low. 

As can be seen in Fig.~\ref{fig_mean_ionization}, the mean ionization increases with temperature, except in the low-temperature, high-density regime, which corresponds to the usual thermal ionization phenomenon. In the low-density, high-temperature regime, the mean ionization decreases with matter density, reflecting the recombination phenomenon. These are common features, also observed in ideal plasmas models. In the high-density regime, above solid density, mean ionzation increases with density, which can be interpreted as a pressure ionization phenomenon.

\begin{figure}[t]
\centerline{\includegraphics[width=8cm]{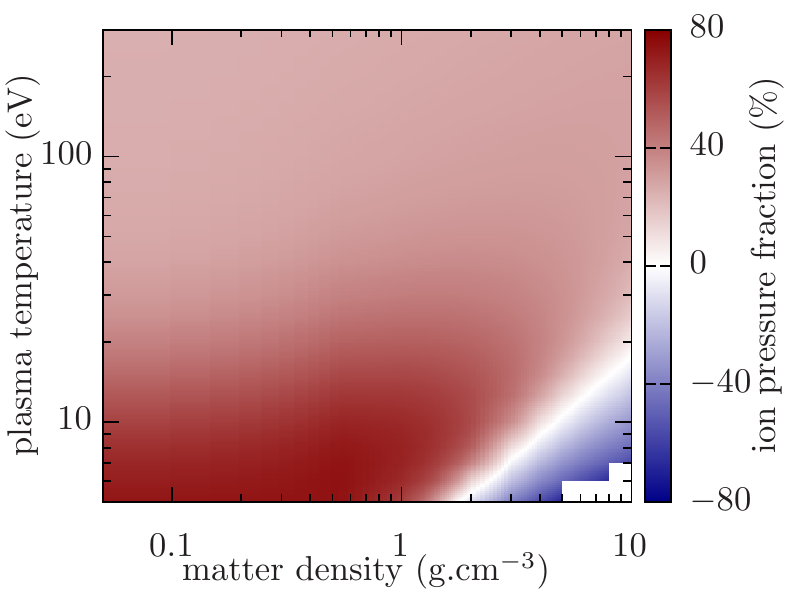}}
\caption{Map of the relative ion fluid contribution to the pressure, for Lithium. In the lower right (blue) region, the excess pressure exceeds the ideal gas contribution in absolute value, leading to a negative ion contribution to the pressure.
\label{fig_ion_contribution_pressure}}
\end{figure}
\begin{figure}[t]
\centerline{\includegraphics[width=8cm]{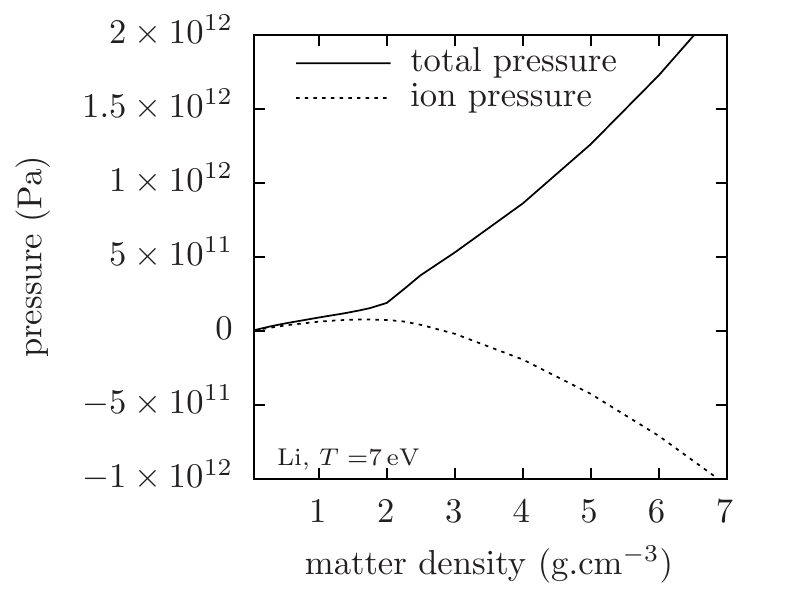}}
\caption{Total and ion-fluid pressures, for Lithium at temperature 7\,eV. The ion contribution to the pressure becomes negative at about 3\,g.cm$^{-3}$.
\label{fig_ion_contribution_pressure_7ev}}
\end{figure}

An interesting feature of the VAMPIRES model is that it provides us with a notion of ion. It also allows us to express thermodynamic quantities as the sum of an ion contribution and a free-electron contribution. Figure \ref{fig_ion_contribution_pressure} displays the fraction of pressure due to the ion fluid (ideal-gas contribution and excess pressure). At high temperature, the ion contribution tends to $25\%$, as is expected for a fully ionized, Lithium ideal plasma. In the case of Lithium, the ion contribution to the pressure is non-negligible everywhere. The line on which this contribution falls to zero indeed corresponds to the transition between positive and negative contribution to the pressure.

In general, negative pressure is a characteristic feature of liquid state. In the present case, it is the pressure of the ion fluid which becomes negative, the total pressure of the plasma remaining positive (see, for instance, Fig.~\ref{fig_ion_contribution_pressure_7ev}). Beyond this qualitative indication of a liquid-gas transition, the validity of our model at these conditions may be questionned, and a clear interpretation of this negative pressure region may be beyond the scope of our approach.

\begin{figure}[t]
\centerline{\includegraphics[width=8cm]{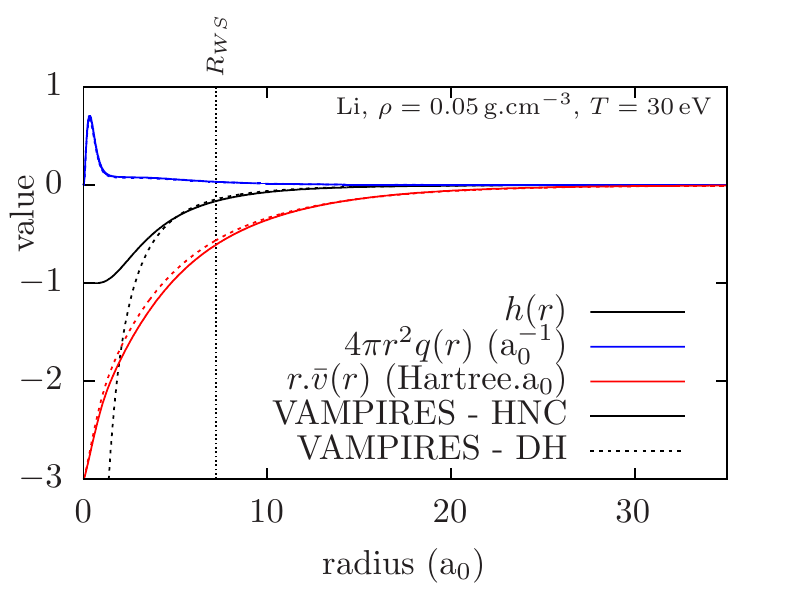}}
\caption{Results from the VAMPIRES model for Lithium at matter density of 0.05\,g.cm$^{-3}$ and temperature of 30\,eV. Plots of the radial correlation function $h(r)$, radial electron density $4\pi r^2 q(r)$, and electron self-consistent potential $r.\bar{v}(r)$, where $\bar{v}(r)=v_\text{el}(r)+v_\text{xc}(n_0+q(r))-v_\text{xc}(n_0)$. Results are presented for both the HNC (solid lines) and DH (dashed lines) versions of the model, which are in good agreement at these conditions. Mean ionization from the HNC version is 2.01, whereas that from the DH version is 2.07.
\label{fig_h_q_uel_005g_30ev}}
\end{figure}

In order to better understand the following analysis, it is useful to define an ``effective coupling parameter'' $\Gamma_\text{eff}=|\beta W^\text{i\,approx}|$ for the ion fluid. Such a number may serve to estimate the role of interactions in a classical fluid with arbitrary potential, just as the usal coupling parameter $\Gamma\equiv\beta e^2Z^{*\,2}/R_\text{WS}$ in the particular case of the OCP. More precisely, the $\Gamma_\text{eff}$ parameter tends to $\sqrt{3}\Gamma^{3/2}/2$ in the Debye-H\"{u}ckel limit of an OCP, and takes values close to $\Gamma$ in the moderate-to-strong coupling regime.

Let us first focus on a case of relatively low density: 0.05\,g.cm$^{-3}$ and moderate temperature: 30\,eV. This case is the object of Fig.~\ref{fig_h_q_uel_005g_30ev}. Effective coupling for this case is low: $\Gamma_\text{eff}=0.28$. In the present case, the HNC and DH versions of the VAMPIRES model yield similar results. Despite the singular behavior of $h(r)$ in the vicinity of $r=0$ in the DH version, the electron self-consistent potential and electron-cloud density are close to those from the HNC version. It is worth mentioning the peak in $q(r)$ close to zero, which mostly corresponds to the bound electrons of the ion. We also see in the $q(r)$ function, a longer-ranged tail of displaced electrons, which is related to the DH-like decay of the electron effective potential and ion-ion correlation function $h(r)$. 

In the VAMPIRES model, as density decreases (and as temperature increases), the effective ion coupling decreases. The radial correlation function $h(r)$ then tends to zero. In view of Eq.~\eqref{eq_def_v_el}, the effective potential seen by the electrons gradually becomes that of an isolated ion, as the Debye screening length tends to infinity. We are then left with an ideal gas of isolated ions, with an ideal gas of free electrons that constitutes a neutralizing background.

\begin{figure}[t]
\centerline{\includegraphics[width=8cm]{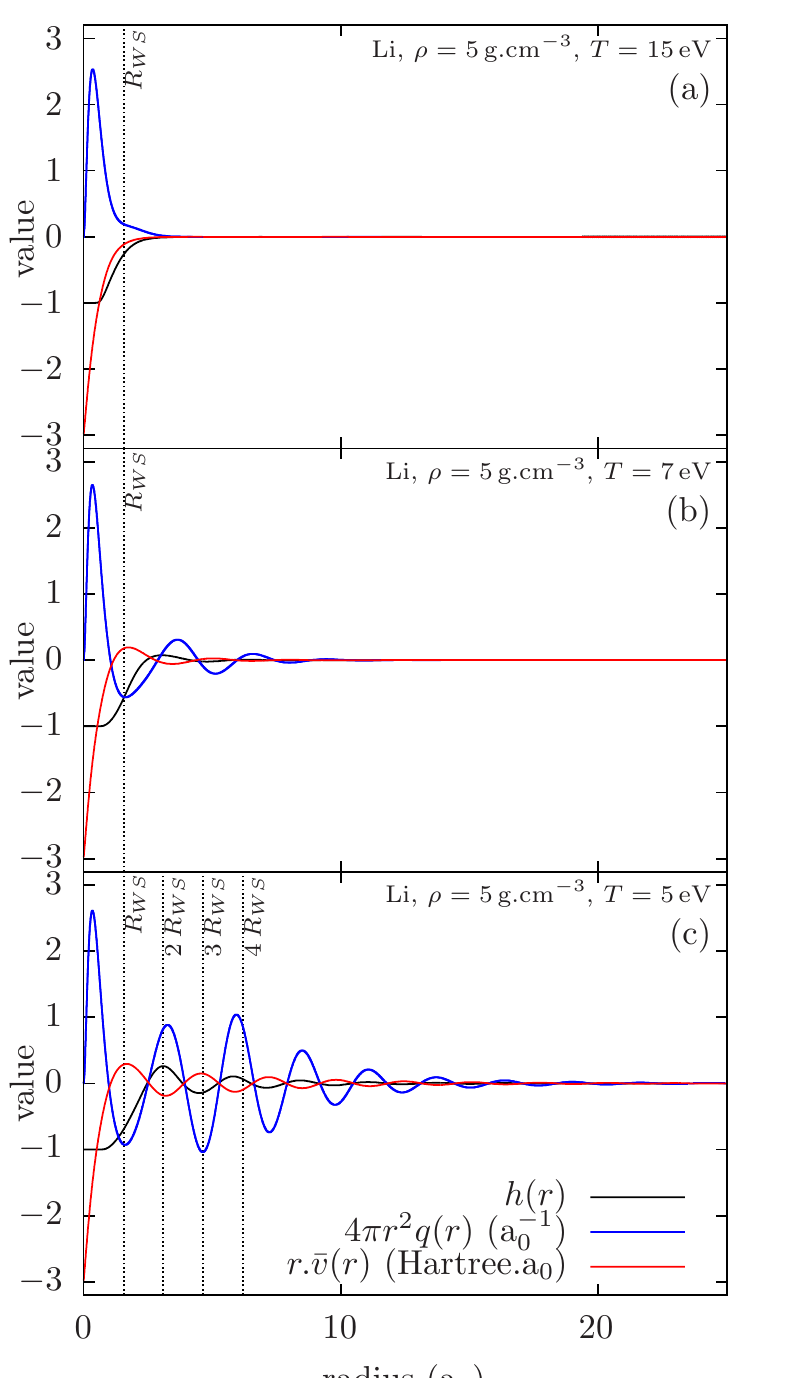}}
\caption{Results from the VAMPIRES model for Lithium at matter density of 5\,g.cm$^{-3}$ and temperatures of 15\,eV, 7\,eV, and 5\,eV. Plots of the radial correlation function $h(r)$, radial electron density $4\pi r^2 q(r)$, and electron self-consistent potential $r.\bar{v}(r)$, where $\bar{v}(r)=v_\text{el}(r)+v_\text{xc}(n_0+q(r))-v_\text{xc}(n_0)$
\label{fig_h_q_uel_5g_5ev}}
\end{figure}

Let us now consider a case of moderate effective coupling: 5\,g.cm$^{-3}$ matter density and temperature of 15\,eV (see Fig.~\ref{fig_h_q_uel_5g_5ev}a). In this case, the effective coupling is $\Gamma_\text{eff}=1.25$. As Debye screening length is decreased, the correlation function $h(r)$ takes a smooth cavity-like shape. Intuitively, we expect such a moderate-coupling case to be among the most similar to what would be obtained from cavity-based model. Due to the smooth rise in $h(r)$, we may still have significant electron polarization outside the WS sphere. However, the Debye length being smaller, this displaced-electron tail is shorter-ranged than in cases of lower coupling.

A case of stronger coupling is that of Fig.~\ref{fig_h_q_uel_5g_5ev}b: 5\,g.cm$^{-3}$ matter density and temperature of 7\,eV. The corresponding effective coupling is high: $\Gamma_\text{eff}=7.15$ and we can see the usual oscillating features of the ion-fluid correlation function $h(r)$, typical of liquid-like behavior. In the VAMPIRES model, these features are convolved with the intra-ion potential $v_\text{intra}(r)$ to build the effective electron potential (see Eq.~\eqref{eq_def_v_el}). In first approximation, it is as if the features of $n_\text{$\nu$}h(r)$ were affected with an effective ion charge $Z^*$. This results in potential wells in phase with the peaks of $h(r)$. On the contrary, between these wells, the  potential overlap of electron clouds leads to repulsive regions. Consequently, electrons are displaced not only towards the central ion but also towards the peaks of $h(r)$, and pushed away from the repulsive region, avoiding too much overlap between electron clouds.

A case of even-stronger coupling ($\Gamma_\text{eff}=13.2$) is showed in Fig.~\ref{fig_h_q_uel_5g_5ev}c: 5\,g.cm$^{-3}$ matter density and temperature of 5\,eV. In such an extreme case, the first peaks in $h(r)$ tend to locate at $2\,R_\text{WS}$ and $4\,R_\text{WS}$, respectively, and may be interpreted as the gradual build up of a first- and second- nearest-neighboring ions. Even if the behavior, explained in previous paragraph, seems qualitatively meaningful, the validity of our main hypotheses regarding the plasma electron density, Eq.~\eqref{eq_approx_average_atom} and the cluster expansion, is of course questionable at these conditions.

Moreover, such strong perturbations of the electron-cloud density, and self-consistent potential, quickly lead to numerical issues, which explain why we did not succeed in performing the computation at higher density for the 5\,eV temperature, and at lower temperature for matter density of 5\,g.cm$^{-3}$.

\section{Comparisons with the VAAQP and INFERNO models}

In the following, we compare the results from the VAMPIRES model to results from two other atom-in-plasma model: VAAQP \cite{Blenski07a,Blenski07b,Piron11}  and INFERNO \cite{Liberman79,Liberman82}. Both of these models allow the calculation of the ion electronic structure through a common quantum formalism for bound and free electrons. Both of these models account for the surrounding plasma using the notion of a Wigner-Seitz (WS) ``cavity''.

The INFERNO model may be considered as an ``ion-in-cell'' model since it is based on the neutrality of the WS sphere, in which the surrounding ions do not enter. Outside the WS sphere, Liberman suggests than the surrounding is modeled as a constant-density jellium whose density is given by the electron chemical potential stemming from the neutrality condition. The discontinuity of the electron density at the WS radius has a strong impact on the thermodynamical consistency of the model, and in particular makes impossible for it to fulfill the virial theorem.

\begin{figure*}[h]
\centerline{
\includegraphics[width=6.5cm]{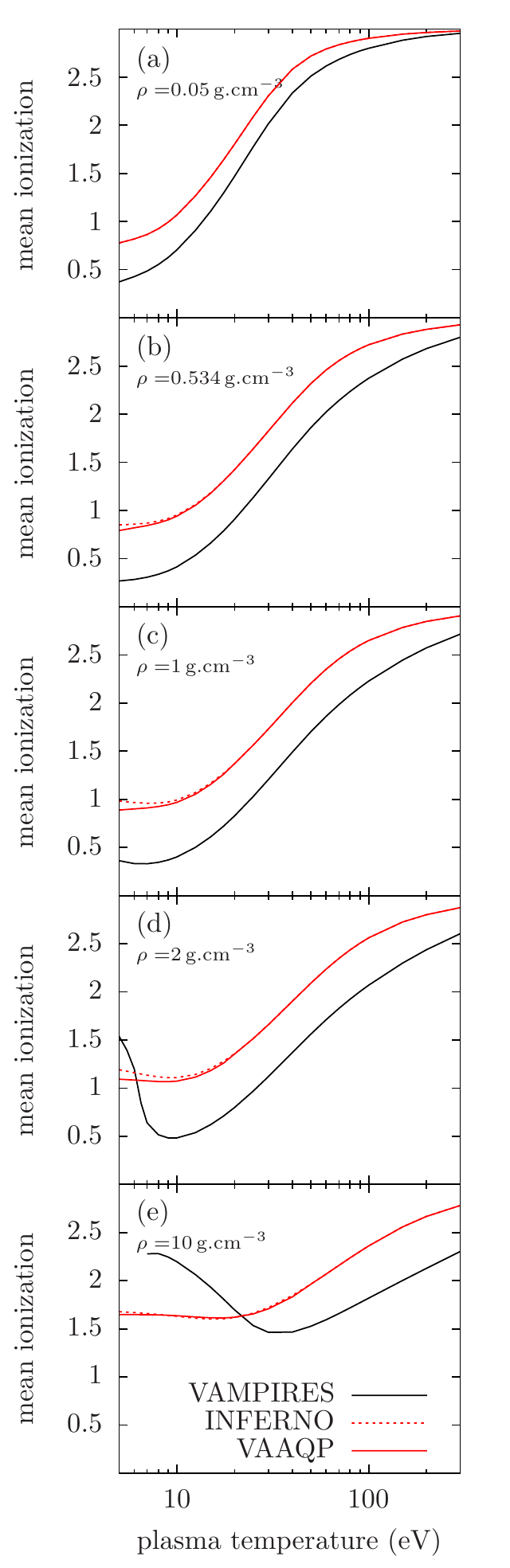}
\includegraphics[width=6.5cm]{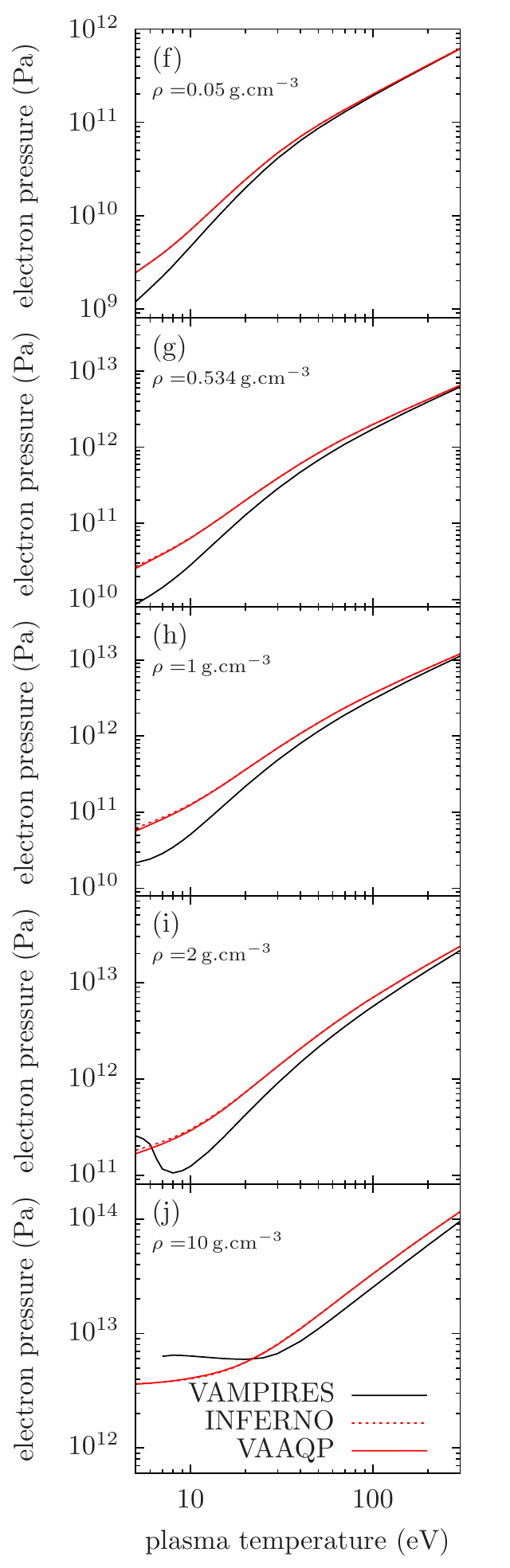}
\includegraphics[width=6.5cm]{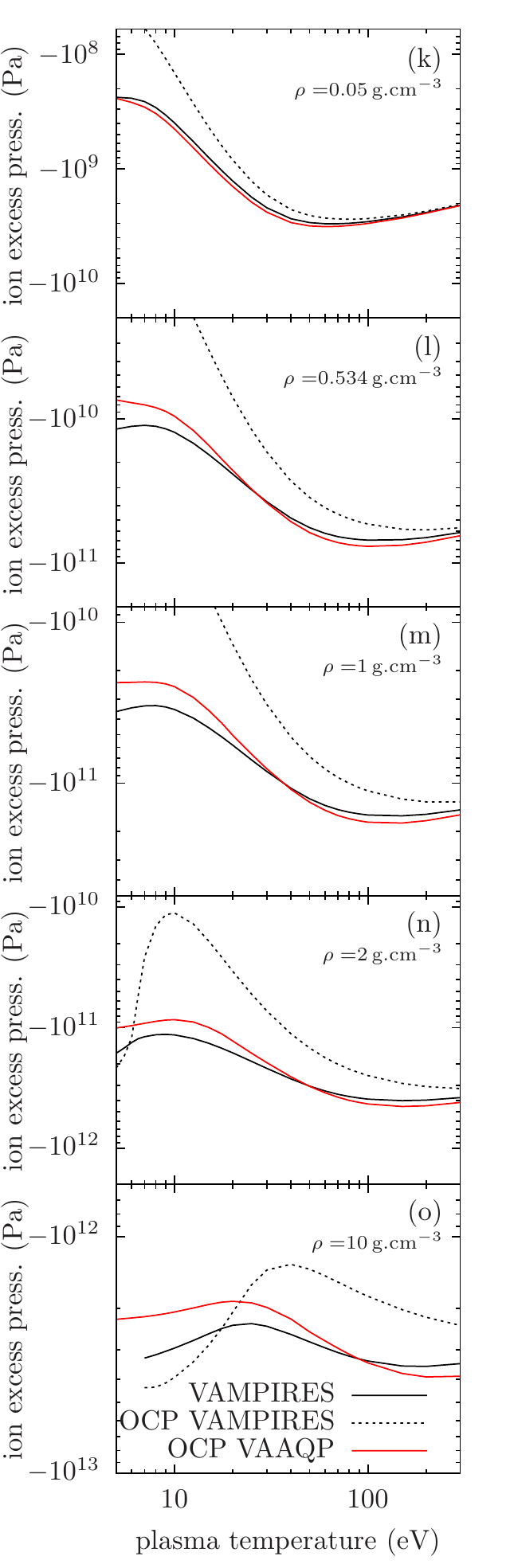}
}
\caption{Comparisons between results from the VAMPIRES model and results from the INFERNO and VAAQP models, on the mean ionization (a--e) and electron pressure (f--j). Comparison between results from the VAMPIRES model and an OCP model with mean ionization taken either from VAMPIRES or from VAAQP, on the ion excess pressure (k--o). Isochoric curves of Lithium, for 5 values of matter density, ranging from $1/10$ solid density to $20$ times the solid density.
\label{fig_comparisons_VAAQP}}
\end{figure*}	
\clearpage

The VAAQP model solves the problem of thermodynamical consistency by requiring the neutrality in the whole space rather than in the sole WS sphere. The WS ``cavity'' however remains a cornerstone of the model, and appears as a region in which surrounding ions do not enter. In a large density-temperature domain, the results from VAAQP were showed to be close to those from the INFERNO approach \cite{Piron11,Piron11b}. In particular, as it can be seen on Fig.~\ref{fig_comparisons_VAAQP}, for the domain of interest in the present study, both approaches mostly agree.

As appears from Eq.~\eqref{eq_pressure_formula}, the pressure of the plasma in the VAMPIRES model may be written as a sum of a contribution from the ion fluid and a contribution from free electrons. The latter contribution is formally similar to the pressure formula obtained from the VAAQP model (see Eq.~98 of \cite{Blenski07b}). The only difference in the formula is the presence of the value of the electrostatic potential at the WS radius in the VAAQP pressure. We recall however that the free-electron density $n_0$ may be different from one model to the other, because it is obtained from a different set of model equations. It is then of interest to compare the mean ionization obtained from these models, as well as to compare the pressure from VAAQP to the free-electron pressure from the VAMPIRES model. Such comparisons are the object of fig~\ref{fig_comparisons_VAAQP}a--j.

Besides, the ion-fluid contribution to the pressure of the VAMPIRES model may be compared to the pressure of a OCP using the mean ionization either from  the VAMPIRES model (OCP-VAMPIRES) or from VAAQP (OCP-VAAQP) to set the ion coupling parameter. The use of a OCP model to supplement the electron pressure found from VAAQP or INFERNO models is a common practice in equation-of-state calculations (see, for instance \cite{Piron11,Piron11b}). Comparisons between the ion-fluid excess pressure from the VAMPIRES model and the excess pressure from the OCP-VAMPIRES and OCP-VAAQP are shown in fig~\ref{fig_comparisons_VAAQP}k--o.

As can be seen in fig~\ref{fig_comparisons_VAAQP}a--e, the mean ionization from the VAMPIRES model is in most conditions lower than that from VAAQP or INFERNO. However, the VAMPIRES model leads to a higher mean ionization in the region of strong pressure ionization, at low temperatures. The picture for the electron pressure comparison (fig~\ref{fig_comparisons_VAAQP}f--j) directly follows from what is seen on the mean ionization.

In Liberman's INFERNO model \cite{Liberman79,Liberman82}, the condition which sets the electron chemical potential $\mu$ or, equivalently, the mean ionization, is the neutrality of the WS sphere. This implies that the effective electrostatic potential seen by the electrons is strictly zero at the WS radius. Being calculated quantum-mechanically, the electron density at the WS radius is in general different from $n_0=n(\mu,T)$, and the the self-consistent potential at the WS radius is given by the $v_\text{xc}(n(r))-v_\text{xc}(n_0)$ contribution, which is usually quite small.

In the VAAQP model \cite{Blenski07a,Blenski07b,Piron11}, the neutrality is required in the whole space and not in the WS sphere. However, the variational calculation results in a condition of cancellation for the integral of the effective electrostatic potential seen by the electrons outside the WS sphere. Even being non-zero, the values of the electrostatic potential at the WS radius are constrained by this condition to be rather small, of the order of the amplitude of the Friedel oscillations. This explains why the VAAQP model often yields results very similar to those of INFERNO, and in substance mostly offers a thermodynamically-sound basis for an INFERNO-like model (see, for instance \cite{Piron11,Piron11b}).  

At a given temperature, the mean ionization stemming from these models, as well as from the VAMPIRES model, is mainly connected to the range of the effective potential seen by the electrons. The shorter the range, the more the bound states are pushed towards the continuum (or even into the continuum under the form of a resonance in the density of states), and the more the chemical potential is pushed towards higher energies in order to fulfill the neutrality condition. 

In that sense, both in VAAQP and in INFERNO, the WS cavity may be seen as a way to force pressure ionization in the model, by restricting the volume occupied by the electronic structure. On the contrary, in the VAMPIRES model, there is no such hypothesis to constrain the range of the effective potential. Its behavior is simply related to the surrounding ions, through the radial correlation function.

\begin{figure}[h]
\centerline{
\includegraphics[width=8cm]{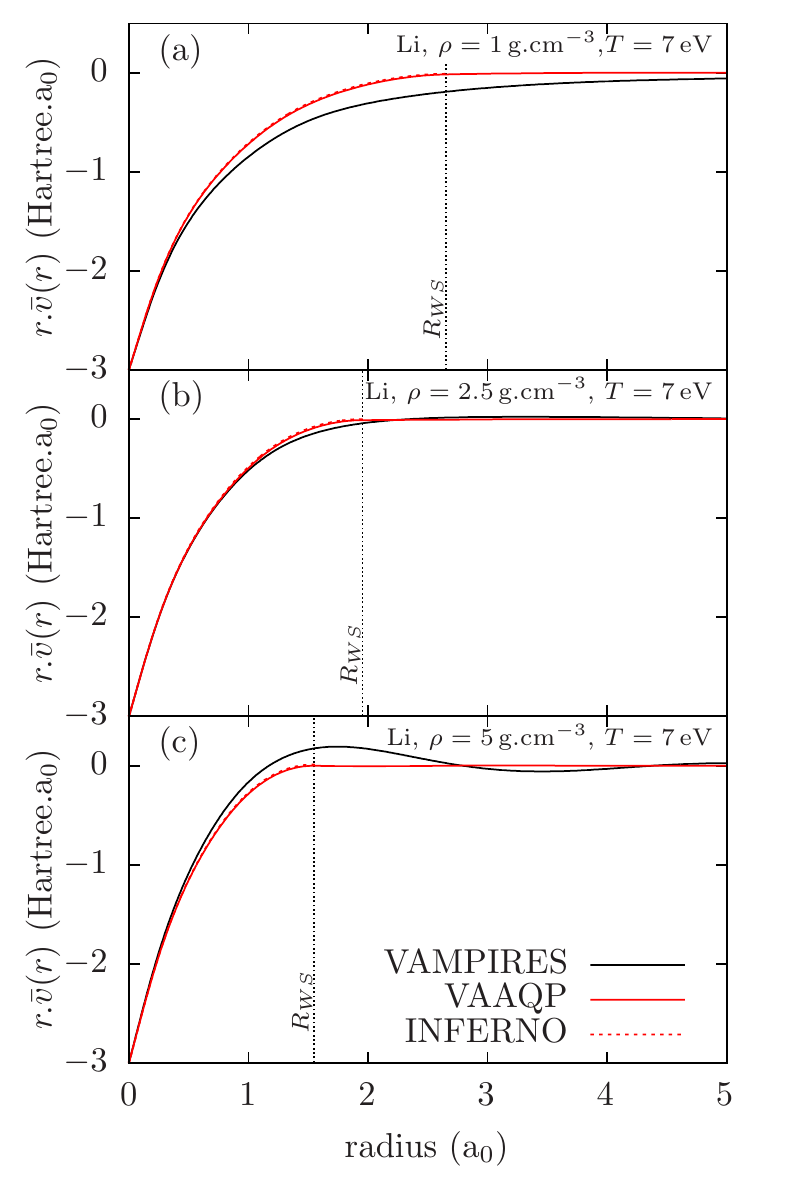}
}
\caption{Comparisons between of the electron self-consistent potential from the VAMPIRES, INFERNO and VAAQP models, across the pressure-ionization front of the 7-eV-isotherm. Chosen matter densities are respectively 1\,g.cm$^{-3}$, 2.5\,g.cm$^{-3}$, and 5\,g.cm$^{-3}$.
\label{fig_comparisons_uel}}
\end{figure}	

As can be seen in fig~\ref{fig_comparisons_VAAQP}, for most of the results we present in this paper, the mean ionization from the VAMPIRES model is lower than the mean ionization obtained from either the VAAQP or the INFERNO model. The explanation is the following. In the VAMPIRE model, the elementary ``building block'' of the effective electrostatic potential $v_\text{el}(r)$ is $v_\text{intra}(r)$, which has a Coulomb tail $Z^*/r$. In itself, the sole electron cloud of the central ion does not fully screen the central ion. Exponential decay of $v_\text{el}(r)$ is assured by the normalization of the radial correlation function $h(r)$. For that reason, the range of $v_\text{el}(r)$ is deeply related to the decay of $h(r)$. As soon as the effective ion-fluid coupling is low, $h(r)$ may extend way farther than the WS radius, its typical decay length being the Debye length (see, for instance Fig.~\ref{fig_h_q_uel_005g_30ev}). The effective potential seen by the electron then extends farther than that of VAAQP or INFERNO, which are constrained to be nearly zero at the WS radius (see Fig.~\ref{fig_comparisons_uel}a). In both VAAQP and INFERNO, the latter constraint stems from the cavity hypothesis, whose validity can be questioned for a plasma at low ion coupling. 

A slightly different standpoint would be to consider that in the VAMPIRES model, the electron cloud can extend way farther than the WS sphere, with a long Debye-Huckel-like tail of displaced electrons, which do not contribute to the asymptotic density $n_0$. In VAAQP or INFERNO, the restriction of the effective-potential range would in some sense categorize \textit{de facto} such electrons as free, making them to participate in $n_0$. In that sense, despite being an asymptotic density in all three models, $n_0$ play a slightly different role in the VAMPIRES model, which may explain its lower values at low-to-moderate ion coupling. 

On the contrary, in the case of a stronger coupling such as that of Fig.~\ref{fig_comparisons_uel}c (or the cases of fig~\ref{fig_h_q_uel_5g_5ev}b and c), the radial correlation function $h(r)$ shows correlation peaks. As was said in the previous section, as regards $v_\text{el}(r)$, this results in potential wells at the peaks of $h(r)$ (i.e. around $2 R_\text{WS}$, $4 R_\text{WS}$...) and repulsive regions in between. As a consequence, $v_\text{el}(r)$ has a first zero at a radius lower than the WS radius. It's attraction range is thus even shorter than in the VAAQP or INFERNO model. This explains why VAMPIRES leads to a higher mean ionization than VAAQP or INFERNO in such cases. It is worth noting that in the VAMPIRES model, pressure ionization is truly a consequence of the accounting for the ion surrondings. It does not come from an hypothesis which results in a direct limitation of the effective-potential range.

\begin{figure}[h]
\centerline{
\includegraphics[width=8cm]{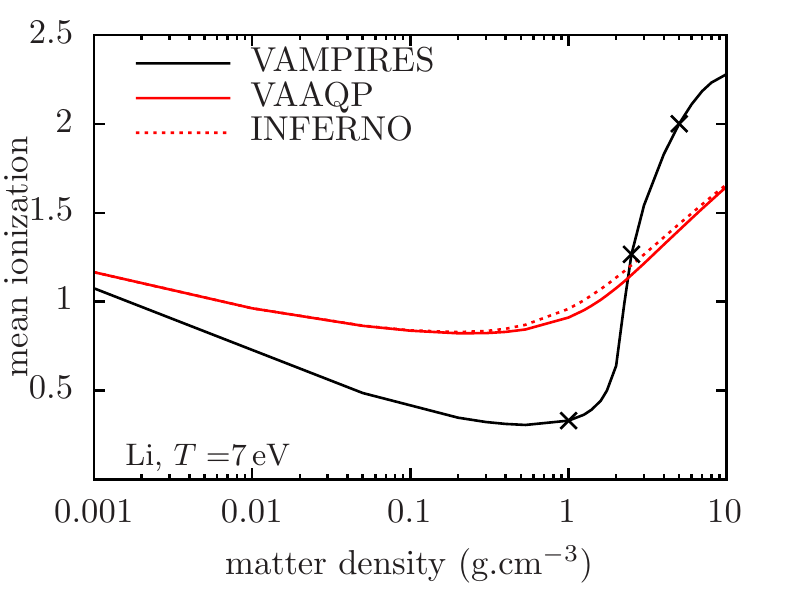}
}
\caption{Comparisons of the mean ionizations from the VAMPIRES, INFERNO and VAAQP models on the 7-eV-isothermal curve, matter density ranging from $1/10$ solid density to $20$ times the solid density. The three crosses correspond to the conditions of Fig.~\ref{fig_comparisons_uel}.
\label{fig_comparisons_VAAQP_7eV}}
\end{figure}	

The fact that pressure ionization results in a sharper rise of the mean ionization as a function of density, in the VAMPIRES model (see Fig.~\ref{fig_comparisons_VAAQP_7eV}) may be seen as a result from the strong modification of $h(r)$ when switching from low to high coupling in the HNC model. In addition to the decrease in the WS radius, the effective potential range switch from a regime in which its range extend farther than the WS radius, to a regime in which it has a significant repulsive feature at the WS radius, and therefore has a zero within a even-shorter range (see Fig.~\ref{fig_comparisons_uel}a,b,c). 

The decrease of mean ionization with temperature which is observed on Fig.~\ref{fig_mean_ionization}, at high densities and low temperatures may be interpreted in the same way. In the strongly-coupled, liquid-like, regime of the ion fluid, an increase in temperature leads to a change in the correlation function that extends the range of the electron effective potential. At these conditions, this effect is stronger than the thermal ionization, which stems from the electron Fermi-Dirac statistics. It thus leads to a decrease in mean ionization.

Figures~\ref{fig_comparisons_VAAQP}k--o shows a comparison of the ion excess pressure stemming from the VAMPIRES model with that stemming from OCP models using either the VAMPIRES mean ionization as an input (OCP-VAMPIRES), or the VAAQP mean ionization (OCP-VAAQP). First, a significant disagreement is obtained between the VAMPIRES results and that from the OCP-VAMPIRES, especially at low densities. This shows the crucial role of accounting for the displaced electrons and their impact on the ion-ion potential. Indeed, the coupling parameter $\Gamma^*\equiv\beta Z^{*\,2}/R_\text{WS}$, is based on $Z^*$, which is only relevant to a region where $h(r)$ is nearly zero. This may explain why integrals such as Eqs~\eqref{eq_int_Wi} or \eqref{eq_virial_fluid} can take very different values in the OCP-VAMPIRES results, even assuming that the OCP-VAMPIRES $h(r)$ is a good approximation to that of VAMPIRES, and especially at low densities.

Interestingly, a much better agreement is obtained with the OCP-VAAQP calculations. This is probably to relate to the previous remark on the slightly different role played by $n_0$ in the two models, which makes $\Gamma^*$ a more relevant approximation in the latter case.

	\section{Conclusions}
In this paper, we propose a model of ion-electron plasma (or nucleus-electron plasma) that accounts for the electronic structure around nuclei (i.e. ion structure) as well as for ion-ion correlations. 
Starting from the problem of the minimization of the plasma free energy, we formulate a series of hypotheses which allows us to approximate this minimization introducing a notion of ion in a plasma. The main hypotheses of this model are:
\begin{enumerate}
\item nuclei are treated as classical indistinguishable particles, whereas all electrons are treated on a same footing, using either a quantum density-functional formalism or a Thomas-Fermi approximation.
\item electronic density is seen as a superposition of a uniform background and spherically-symmetric distributions around each nucleus (system of ions in a plasma) 
\item free energy is approached using a cluster expansion (non-overlapping ions) 
\item resulting ion fluid is modeled through an approximate integral equation.
\end{enumerate}

The model equations are then obtained through the minimization of the approximate free-energy functional. The result is a self-consistent model of the ion electronic structure and ion-ion correlations. In this model, each ion is partially screened by its own electron cloud, the complete screening of the ion resulting from the interaction with its surrounding ions and the free electrons.

The model being obtained from a variational approach, thermodynamical quantities are easily derived. In particular, a simple formula is obtained for the pressure. 
Since the model defines the notion of an ion in a plasma, each thermodynamical quantity may be written as the sum of an electron contribution and an ion contribution, which may open the way to heuristical applications to two-temperature systems. It is eventually shown that the model fulfills the virial theorem. In the present paper, the model is only presented in its average-atom version. 

Numerical applications of the VAMPIRES model to a Lithium plasma at compressions $1/10$ to $20$, and at temperatures ranging from 5\,eV to 300\,eV are showed and discussed. Results from the VAMPIRES model are also compared to two other atom-in-plasma models, based on the Wigner-Seitz cavity.

DFT molecular-dynamics approaches based on a modified ``canonical'' dynamics \cite{Nose84,Hoover85} in principle address the calculation of the canonical equilibrium quantities. Such quantities can be compared to those from the VAMPIRES model. Such comparisons are left for a subsequent study.

Extensions of the VAMPIRES model to a detailed accounting of the ion excited states, as well as to the description of plasma mixtures were already obtained by the authors, and will be the subject of upcoming publications.

	\section*{Acknowledgements}
	The authors would like to thank Bogdan Cichocki (University of Warsaw) for useful discussions at the beginning of this theoretical work. They are also grateful to the LERMA/Sorbonne Universit\'{e}, especially to Andrea Ciardi and Franck Delahaye, who hosted them on many occasions during the late phase of this study.

	\appendix
	
	\section{Notations\label{app_notations}}
	For the sake of shortening the notations, we use the following conventions.
	\begin{align} 
		F\{\underline{w};y_1,y_2,...\}
	\end{align}
	denotes a functional $F$ of a function $w(x_1,x_2,...)$ and variables $y_1, y_2,...$. Functional dependencies are underlined and most-often written without mentioning their own arguments. In this case, the variables on which the function depends are stated in the body of the text.
	\begin{align} 
		F_\text{min}(y_1,y_2,...,Z)=&\underset{\underline{w}}{\text{Min}}\,F\{\underline{w};y_1,y_2,...\}
		\nonumber\\
		&\text{s.\,t. } G\{\underline{w};y_1,y_2,...\}=Z
	\end{align} 
	denotes the minimum of the functional $F\{\underline{w};y_1,y_2,...\}$ with respect to the function $w(x_1,x_2,...)$, subject to the constraint $G\{\underline{w};y_1,y_2,...\}=Z$.
	\begin{align} 
		w_\text{min}(x_1,x_2,...;&y_1,y_2,...,Z)\nonumber\\
		=&\underset{\underline{w}}{\text{Argmin}}\,F\{\underline{w};y_1,y_2,...\}
		\nonumber\\
		&\text{s.\,t. } G\{\underline{w};y_1,y_2,...\}=Z
	\end{align} 
	denotes the value, in the sense of functions, which minimizes the functional while fulfilling the constraint.
	
	In order to shorten the notation of functions dependencies, we will note the nuclei configurations and nuclei classical states, respectively, as follows:
	\begin{align}
		&\vec{R}_1...\vec{R}_{N_\text{$\nu$}}
		\equiv \vec{R}_1,\vec{R}_2...\vec{R}_{N_\text{$\nu$}}\\
		&\vec{R}_1...\vec{P}_{N_\text{$\nu$}}
		\equiv \vec{R}_1,\vec{R}_2...\vec{R}_{N_\text{$\nu$}},
		\vec{P}_1,\vec{P}_2...\vec{P}_{N_\text{$\nu$}}
	\end{align}
	In the same spirit, the integral that corresponds to the sum over nuclei classical states will be written:
	\begin{align}
		&\iint_V\frac{d^3R_1...d^3P_{N_\text{$\nu$}}}{N_\text{$\nu$}!h^{3N_\text{$\nu$}}}
		\nonumber\\
		&\equiv \frac{1}{N_\text{$\nu$}!}\int_V d^3R_1...\int_V d^3R_{N_\text{$\nu$}}
		\int \frac{d^3P_1}{h^3}...\int \frac{d^3P_{N_\text{$\nu$}}}{h^3}
	\end{align}
	
	In order to shorten the equations, function dependencies are sometimes omitted in the equations, in such case, they are mentioned explicitly in the body of the text.
	
	\section{Generalized form of the classical-nuclei approximation\label{app_generalized_zwanzig}}
	The classical limit of the quantum statistical mechanics may be addressed through the Wigner-Kirkwood expansion \cite{Kirkwood33,Kirkwood34}. More specifically the case of a nucleus-electron mixture with only nuclei taken in the classical limit was studied in \cite{Zwanzig57}. In the fully-classical limit for indistinguishable nuclei, it is shown in the latter reference that the partition function $Q(N_\text{$\nu$},V,T)$ of the nucleus-electron system writes:
	\begin{align}
		&Q(N_\text{$\nu$},V,T)\nonumber\\
		&=\iint_V\frac{d^3R_1...d^3P_{N_\text{$\nu$}}}{N_\text{$\nu$}!h^{3N_\text{$\nu$}}}
		\left\{
		\exp\left(-\beta \left( 
		\sum_{j=1}^{N_\text{$\nu$}}\frac{P_j^2}{2m_\text{$\nu$}}
		+F_\text{eq}^\text{e}
		\right) \right)
		\right\}
	\end{align}
	where $F_\text{eq}^\text{e}(\vec{R}_1...\vec{R}_{N_\text{$\nu$}};N_\text{$\nu$},V,T)$ is the free energy of electrons, in the potential generated by nuclei at fixed positions $\vec{R}_1...\vec{R}_{N_\text{$\nu$}}$, including also the nucleus-nucleus interaction energy:
	\begin{align}
		\frac{1}{2}\sum_{i=1}^{N_\text{$\nu$}}\sum_{\substack{j=1\\j\neq i}}^{N_\text{$\nu$}}\frac{Z^2e^2}{|\vec{R}_i-\vec{R}_j|} 
	\end{align}
	
	The corresponding classical canonical distribution, which allows one to perform the canonical averages over the statistical ensemble, is:
	\begin{align}
		w^\text{eq}(\vec{R}_1...\vec{P}_{N_\text{$\nu$}};N_\text{$\nu$},V,T)
		=\frac{e^{-\beta \left( 
				\sum_{j=1}^{N_\text{$\nu$}}\frac{P_j^2}{2m_\text{$\nu$}}
				+F_\text{eq}^\text{e}
				\right)}}{Q(N_\text{$\nu$},V,T)}
				\label{eq_canonical_distribution_w}
	\end{align}
	and the corresponding statistical entropy writes:
	\begin{align}
		S&\left\{w^\text{eq}\right\}\nonumber\\
		=&-k_\text{B}
		\iint_V\frac{d^3R_1...d^3P_{N_\text{$\nu$}}}{N_\text{$\nu$}!h^{3N_\text{$\nu$}}}
		\left\{
		w^\text{eq}\log\left(w^\text{eq}\right)
		\right\}\\
		=&k_\text{B}\beta
		\iint_V\frac{d^3R_1...d^3P_{N_\text{$\nu$}}}{N_\text{$\nu$}!h^{3N_\text{$\nu$}}}
		\left\{
		w^\text{eq} \left( 
		\sum_{j=1}^{N_\text{$\nu$}}\frac{P_j^2}{2m_\text{$\nu$}}
		+F_\text{eq}^\text{e}
		\right)
		\right\}\nonumber\\
		&+k_\text{B}\log\left(Q\right)\label{eq_stat_entropy}
	\end{align}
	The free energy, as it is defined from the partition function, is:
	\begin{align}
		F_\text{eq}(N_\text{$\nu$},V,T)
		=&-\frac{1}{\beta}\log\left(Q\right)
	\end{align}
	From Eq.~\eqref{eq_stat_entropy}, we may thus write the free energy as:
	\begin{align}
		&F_\text{eq}(N_\text{$\nu$},V,T)
		\nonumber\\
		&=\iint_V\frac{d^3R_1...d^3P_{N_\text{$\nu$}}}{N_\text{$\nu$}!h^{3N_\text{$\nu$}}}
		\left\{
		w^\text{eq} \left( 
		\sum_{j=1}^{N_\text{$\nu$}}\frac{P_j^2}{2m_\text{$\nu$}}
		+F_\text{eq}^\text{e}
				\right.\right.\nonumber\\&\hspace{4cm}\left.\left.
		+\frac{1}{\beta}\log\left(w^\text{eq}\right)
		\vphantom{\sum_{j=1}^{N_\text{$\nu$}}}\right)
		\right\}
	\end{align}
	Let us now generalize the latter expression of the free energy to arbitrary statistical distributions $w\left(\vec{R}_1...\vec{P}_{N_\text{$\nu$}}\right)$. We then obtain the following generalized-free-energy functional:
	\begin{align}
		&F\left\{\underline{w};N_\text{$\nu$},V,T\right\}
		\nonumber\\
		&=\iint_V\frac{d^3R_1...d^3P_{N_\text{$\nu$}}}{N_\text{$\nu$}!h^{3N_\text{$\nu$}}}
		\left\{
		w \left( 
		\sum_{j=1}^{N_\text{$\nu$}}\frac{P_j^2}{2m_\text{$\nu$}}
		+F_\text{eq}^\text{e}
		+\frac{1}{\beta}\log\left(w\right)
		\right)
		\right\}
	\end{align}
	It is easy to check that minimizing this generalized-free-energy functional $F$ with respect to the statistical distribution $w$, while requiring its normalization to 1, we recover the classical canonical distribution $w^\text{eq}$, for which $F$ takes the value $F_\text{eq}$. We summarize this variational formulation of the free-energy calculation as follows:
	\begin{align}
		F_\text{eq}&(N_\text{$\nu$},V,T)\nonumber\\
		=&\underset{\underline{w}}{\text{Min}}\,F\left\{\underline{w};N_\text{$\nu$},V,T\right\}
		\nonumber\\
		&\text{s.\,t. } \int_V \frac{d^3R_1...d^3P_{N_\text{$\nu$}}}{N!h^{3N_\text{$\nu$}}}\left\{w(\vec{R}_1...\vec{P}_{N_\text{$\nu$}})\right\}=1
	\end{align}

	\section{Useful expressions and functional derivatives\label{app_functional_derivatives}}
	\subsection{Fluid integral equations from the generalized-free-energy functionals\label{app_deriv_int_equations}}
The expressions for the fluid generalized-free-energy functionals in the HNC and DH models are given in Eqs. \eqref{eq_hnc_free_energy_renorm} and \eqref{eq_dh_free_energy_renorm}, respectively. For the sake of keeping the paper self-contained, we recall below the derivation of the integral equations for these two models, starting from the corresponding free-energy functionals.

The fluid integral equations are obtained by minimizing the generalized-free-energy functionals with respect to the radial correlation function $h(r)$:
\begin{align}
\frac{\delta\bar{A}^\text{i\,approx}}{\delta h(r)}=0
\end{align}
Let us first evaluate the functional derivative that follows:
\begin{align}
D_1\equiv\frac{\delta}{\delta h(r)}\left(
\int \frac{d^3k}{(2\pi)^3}\left\{n_\text{$\nu$} h_k-
\log\left(1+n_\text{$\nu$} h_k\right)
\right\}\right)
\end{align}
which is common to the HNC and DH cases. In view of the definition of the Fourier transform of $h(r)$, Eq.~\eqref{eq_dh_free_energy_renorm} we have:
\begin{align}
\frac{\delta h_\vec{k}}{\delta h(\vec{r})}=e^{i\vec{k}.\vec{r}}
\end{align}
we then have:
\begin{align}
D_1=&
\int \frac{d^3k}{(2\pi)^3}\left\{\left(1-
\frac{1}{1+n_\text{$\nu$} h_k}\right)n_\text{$\nu$} e^{i\vec{k}.\vec{r}}
\right\}\\
=&\int \frac{d^3k}{(2\pi)^3}\left\{n_\text{$\nu$}^2 c_k  e^{i\vec{k}.\vec{r}}
\right\}=n_\text{$\nu$}^2 c(r)
\end{align}
where we \emph{define} $c_k$ as follows:
\begin{align}
c_k\equiv \frac{h_k}{1+n_\text{$\nu$} h_k}
\end{align}
which yields in the direct space:
\begin{align}
c(r) \equiv h(r)- n_\text{$\nu$}\int d^3r'\left\{h(r')c(|\vec{r}-\vec{r}'|)\right\}
\label{eq_oz_appendix}
\end{align}
The latter relation is indeed the Ornstein-Zernicke relation.

For the HNC case, we differentiate the first term of Eq. \eqref{eq_hnc_free_energy_renorm} with respect to $h(r)$, and get:
\begin{align}
\frac{\delta\bar{A}^\text{HNC}}{\delta h(r)}=&
\frac{n_\text{$\nu$}}{2\beta}\left(\beta v(r) + \log\left(h(r)+1\right)-h(r)\right)+\frac{D_1}{2\beta n_\text{$\nu$}}
\end{align}
and the minimization condition gives:
\begin{align}
c(r)=-\beta v(r) - \log\left(h(r)+1\right)+h(r)
\end{align}
which is the HNC closure relation.

For the DH case, we differentiate the first term of Eq. \eqref{eq_dh_free_energy_renorm} with respect to $h(r)$, and get:
\begin{align}
\frac{\delta\bar{A}^\text{DH}}{\delta h(r)}=&
\frac{n_\text{$\nu$}}{2\beta}\beta v(r)+\frac{D_1}{2\beta n_\text{$\nu$}}
\end{align}
and the minimization condition gives:
\begin{align}
c(r)=-\beta v(r)
\end{align}
which is the DH closure relation, leading to the DH integral equation when it is used in Eq.~\eqref{eq_oz_appendix}.

	\subsection{Expressions of $\Delta F_1^{0}$ in the quantum and Thomas-Fermi cases\label{app_expr_dF10}}
In our approach, $\Delta F_1^{0}$ is the one-center contribution to the kinetic-entropic term of the free energy. Its expression is slightly different from the DFT free-energy expression for finite systems. According to the one-center cluster expansion, we have the following structure for $\Delta F_1^{0}$:
\begin{align}
&\Delta F_1^{0}\left\{\underline{q},n_0;T\right\}\nonumber\\
&=\int d^3r
\left\{f_1^{0}\{\underline{n}(r')=n_0+q(r');\vec{r};T\} - f_0^{0}(n_0;T)\right\}
\label{free_energy_difference_integral}
\end{align}
It is a spatial integral of the difference between two electron free-energy-densities, each corresponding to an infinite medium. The first is free-energy-density of an inhomogeneous electron gas of density $n_0+q(r)$. The second is free-energy-density of a homogeneous electron gas of density $n_0$. Let us denote by $\bar{v}\left\{\underline{q},n_0;r;T\right\}$ the effective (or trial) potential leading to the inhomogeneous density $n_0+q(r)$. 

In the quantum-mechanical approach $f_1^{0}\{\underline{n};\vec{r};T\}$ is a non-local functional of the density $n(r)$.
\begin{align}
f_1^{0\,\text{QM}}&\{\underline{n}(r')=n_0+q(r');\vec{r};T\} =\nonumber\\
2\sum_j & \left(
f_\text{FD}(\varepsilon_j,\mu_0,T)
\psi_j^*(\vec{r})\frac{-\hbar^2}{2m_\text{e}}\nabla^2_\vec{r}\psi_j(\vec{r})
\right.\nonumber\\&\left.
-\frac{1}{\beta} 
s\left(f_\text{FD}(\varepsilon_j,\mu_0,T)\right)\left|\psi_j(\vec{r})\right|^2\right)
\nonumber\\
+2\int& \frac{d^3p}{h^3}\left\{
f_\text{FD}(\varepsilon_p,\mu_0,T)\psi_\vec{p}^*(\vec{r})\frac{-\hbar^2}{2m_\text{e}}\nabla^2_\vec{r}\psi_\vec{p}(\vec{r})
\right.
\nonumber\\&
\left.
-\frac{1}{\beta}s\left(f_\text{FD}(\varepsilon_p,\mu_0,T)\right)|\psi_\vec{p}(\vec{r})|^{2}\right\}
\label{free_energy_density_QM}
\end{align}
where $\varepsilon_p\equiv p^2/(2m_\text{e})$, where the entropy function $s(f)$ is defined as follows:
\begin{align}
s(f)= -(&f\log(f)+(1-f)\log(1-f))
\label{entropy_s}
\end{align}
and where the wave-functions $\psi_j(\vec{r})$ and $\psi_\vec{p}(\vec{r})$ are indeed functionals of the effective potential $\bar{v}(r)=\bar{v}\left\{\underline{q},n_0;r;T\right\}$, and defined through the Schr\"{o}dinger Eq.~\eqref{eq_dft_schrodinger_1electron}.

Eq.~\eqref{free_energy_density_QM} can be extended straightforwardly to an electron system described using the Dirac equation instead of the Schr\"{o}dinger equation (see, for instance \cite{Piron11}).

The Thomas-Fermi approach is based on a local approximation to the kinetic-entropic free-energy. We then have for $f_1^{0}\{\underline{n};\vec{r};T\}$ a function of the local value $n(r)$. It is in fact $f_0(n,T)$, the same function as the free energy per unit volume of an homogeneous plasma of density $n$. Its expression as a function of the effective potential $\bar{v}(r)$ may be written as:
\begin{align}
	f_1^{0\,\text{TF}}&\{\underline{n}(r')=n_0+q(r';\vec{r});T\} =f_0(n_0+q(r),T)\nonumber\\
	=&2\int \frac{d^3p}{h^3}\Big\{
	f_\text{FD}\big(\varepsilon_p +\bar{v}\{\underline{q},n_0;r;T\},\mu_0,T \big)\varepsilon_p
	\nonumber\\
	&-\frac{1}{\beta}s\big(f_\text{FD}(\varepsilon_p +\bar{v}\{\underline{q},n_0;r;T\},\mu_0,T)\big)\Big\}
\end{align}

	\subsection{Proof of the Eqs. \eqref{eq_dF10_dq} and \eqref{eq_dF10dn0} }
We first simplify the notation in Eq.\eqref{eq_dF10_dq} writing $\bar{v}(r)$ instead of $\bar{v}\{\underline{q},n_0;r;T\}$ and $f_a$ instead of$f_\text{FD}(\varepsilon_a,\mu_0,T)$, where $a$ may be either a discrete $j$ value or a continuous $p$ value. 

From Eqs~\eqref{free_energy_difference_integral}, \eqref{free_energy_density_QM}, using the Scr\"{odinger} equation Eq.~\eqref{eq_dft_schrodinger_1electron} and the normalization of the wave-functions, we have:
\begin{align}
\Delta F_1^{0\,\text{QM}}=&
2\sum_j \left(
f_j\varepsilon_j
-\frac{1}{\beta} s\left(f_j\right)
\right)
\nonumber\\
&+2\int \frac{d^3p}{h^3}\left\{
\left( f_p\varepsilon_p-\frac{1}{\beta}s\left(f_p\right) \right) C_\vec{p}\right\}
\nonumber\\
&-\int d^3r\left\{\left(n_0+q(r)\right)\bar{v}(r)\right\}
\label{eq_expr_dF10_appendix}
\end{align}
where we define:
\begin{align}
C_\vec{p}\equiv \int d^3r\left\{|\psi_\vec{p}(\vec{r})|^{2} - 1\right\}
\end{align}
which is a functional of the effective potential $\bar{v}(r)=\bar{v}\left\{\underline{q},n_0;r;T\right\}$.

From the the static, first-order perturbation theory (see, for instance \cite{Schiff}), we have:
\begin{align}
&\frac{\delta\varepsilon_j}{\delta \bar{v}(r)} = |\psi_j(\vec{r})|^{2}\label{eq_perturb_eigenvalue}
\end{align}
\begin{align}
\frac{\delta \psi_j(\vec{r})}{\delta \bar{v}(\vec{r}')} 
=& \sum_{j'\neq j} \psi_{j'}(\vec{r})\frac{ \psi_{j'}^*(\vec{r}')  \psi_{j}(\vec{r}') }{\varepsilon_j-\varepsilon_{j'}}
\nonumber\\
&+\int \frac{d^3p'}{h^3}\left\{\psi_{\vec{p}'}(\vec{r})  \frac{ \psi_{\vec{p}'}^*(\vec{r}')  \psi_{j}(\vec{r}') }{\varepsilon_j-\varepsilon_{p'}}\right\}
\\
\frac{\delta \psi_{\vec{p}}(\vec{r})}{\delta \bar{v}(\vec{r}')} 
=& \sum_{j} \psi_{j'}(\vec{r})\frac{ \psi_{j'}^*(\vec{r}')  \psi_{\vec{p}}(\vec{r}') }{\varepsilon_p-\varepsilon_{j'}}
\nonumber\\
&+\int \frac{d^3p'}{h^3}\left\{\psi_{\vec{p}'}(\vec{r})  \frac{ \psi_{\vec{p}'}^*(\vec{r}')  \psi_{\vec{p}}(\vec{r}') }{\varepsilon_p-\varepsilon_{p'}}\right\}
\end{align}
Starting from the latter equation, we may then show the following relations for any function $g_p$:
\begin{align}
&\frac{\delta }{\delta \bar{v}(\vec{r}')} \int \frac{d^3p}{h^3} \left\{g_p C_\vec{p}\right\}
\nonumber\\
&=\int \frac{d^3p}{h^3} \left\{g_p \int d^3r\frac{\delta |\psi_{\vec{p}}(\vec{r})|^2}{\delta \bar{v}(\vec{r}')} \right\}
\\
&=\int\frac{d^3p}{h^3}\int\frac{d^3p'}{h^3}\int d^3r \left\{g_p \psi_{\vec{p}}^*(\vec{r})\psi_{\vec{p}'}(\vec{r})  \frac{ \psi_{\vec{p}'}^*(\vec{r}')  \psi_{\vec{p}}(\vec{r}') }{\varepsilon_p-\varepsilon_{p'}} \right\}
\nonumber\\
&+\int\frac{d^3p}{h^3}\int\frac{d^3p'}{h^3}\int d^3r \left\{g_p \psi_{\vec{p}}(\vec{r})\psi_{\vec{p}'}^*(\vec{r})  \frac{ \psi_{\vec{p}'}(\vec{r}')  \psi_{\vec{p}}^*(\vec{r}') }{\varepsilon_p-\varepsilon_{p'}} \right\}
\\
&=\int\frac{d^3p}{h^3}\int\frac{d^3p'}{h^3} \left\{\frac{g_p-g_{p'}}{\varepsilon_p-\varepsilon_{p'}}\vphantom{\int}
\right.\nonumber\\&\hspace{1cm}\times\left.
 \int d^3r  \left\{\psi_{\vec{p}}^*(\vec{r})\psi_{\vec{p}'}(\vec{r})\psi_{\vec{p}'}^*(\vec{r}')  \psi_{\vec{p}}(\vec{r}') \right\} \right\}
\\
&=\int_0^\infty \frac{dp}{h^3}\left\{p^2\frac{\partial g_p}{\partial \varepsilon_p}\int_{4\pi}d\Omega_{\vec{p}}\left\{ |\psi_{\vec{p}}(\vec{r})|^{2}\right\}\right\}  
\label{eq_deriv_Cp}
\end{align}
where $\Omega_{\vec{p}}$ is the solid angle corresponding to $\vec{p}$.

%

We now perform the differentiation of Eq.~\eqref{eq_expr_dF10_appendix} with respect to $q(r)$
\begin{align}
&\frac{\delta \Delta F_1^{0\,\text{QM}}}{\delta q(r)}\nonumber\\
&=\int d^3r'\left\{\frac{\delta\bar{v}(r')}{\delta q(r)} \left(
2\sum_j\frac{\partial}{\partial\varepsilon_j}\left(f_j\varepsilon_j-\frac{1}{\beta}s\left(f_j\right)\right)
\frac{\delta \varepsilon_j}{\delta\bar{v}(r')}
\right.\right.
\nonumber\\&\hphantom{=\int d^3r'}
+2\frac{\delta}{\delta \bar{v}(r')}\int \frac{d^3p}{h^3}\left\{
\left( f_p\varepsilon_p-\frac{1}{\beta}s\left(f_p\right) \right) C_\vec{p}\right\}
\nonumber\\&\hphantom{=\int d^3r'}\left.\left.
-\left(n_0+q(r')\right)\vphantom{\left(\sum_j\right)}\right)\right\}-\bar{v}(r)
\end{align}
	
From Eq.~\eqref{entropy_s} and the expression of the Fermi-Dirac distribution, we have:
\begin{align}
&\frac{\partial}{\partial\varepsilon_a}\left(f_a\varepsilon_a-\frac{1}{\beta}s\left(f_a\right)\right)
=\mu_0\frac{\partial f_a}{\partial\varepsilon_a}+f_a
\label{eq_deriv_s_f}
\end{align}

Using Eqs~\eqref{eq_perturb_eigenvalue}, \eqref{eq_deriv_Cp}, \eqref{eq_deriv_s_f} we obtain:
\begin{align}
&\frac{\delta \Delta F_1^{0\,\text{QM}}}{\delta q(r)}\nonumber\\
&=\int d^3r'\left\{\frac{\delta\bar{v}(r')}{\delta q(r)} \left(
2\sum_j\left(\mu_0\frac{\partial f_j}{\partial\varepsilon_j}+f_j\right)
|\psi_j(\vec{r}')|^{2}
\right.\right.
\nonumber\\&
+2\int_0^\infty \frac{dp}{h^3}\left\{p^2 \left(\mu_0\frac{\partial f_p}{\partial\varepsilon_p}+f_p\right) \int_{4\pi}d\Omega_{\vec{p}}\left\{ |\psi_{\vec{p}}(\vec{r}')|^{2}\right\}\right\}
\nonumber\\&\hphantom{=\int d^3r'}\left.\left.
-\left(n_0+q(r')\right)\vphantom{\left(\sum_j\right)}\right)\right\}-\bar{v}(r)
\\
&=\int d^3r'\left\{\mu_0\frac{\delta\bar{v}(r')}{\delta q(r)} \left(
2\sum_j\frac{\partial f_j}{\partial\varepsilon_j}
|\psi_j(\vec{r}')|^{2}
\right.\right.
\nonumber\\&\left.\left.
+2\int_0^\infty \frac{dp}{h^3}\left\{p^2 \frac{\partial f_p}{\partial\varepsilon_p} \int_{4\pi}d\Omega_{\vec{p}}\left\{ |\psi_{\vec{p}}(\vec{r}')|^{2}\right\}\right\}\vphantom{\left(\sum_j\right)}\right)\right\}
-\bar{v}(r)
\end{align}
Finally, using again Eq.~\eqref{eq_perturb_eigenvalue}, \eqref{eq_deriv_Cp}, in order to go back to derivatives with respect to $v(r)$, we obtain:
\begin{align}
&\frac{\delta \Delta F_1^{0\,\text{QM}}}{\delta q(r)}
=\mu_0\int d^3r'd^3r''\left\{\frac{\delta\bar{v}(r')}{\delta q(r)}\frac{\delta q(r'')}{\delta \bar{v}(r')}\right\}
\\
&=\mu_0-\bar{v}(r)
\end{align}	
	
We now calculate the derivative of $\Delta F_1^0$ with respect to $n_0$ at fixed $q(\vec{r})$. 

From equation \eqref{free_energy_difference_integral}, we know that $\Delta F_1^0$ can be written as a functional of $n(r) = n_0+q(r)$, and $n_0$. Let us define:
\begin{align}
\Delta \tilde{F}_1^0\left\{\underline{n}(r)=n_0+q(r),n_0\right\}\equiv \Delta F_1^0\left\{\underline{q}(r),n_0\right\}
\label{eq_dF1_dF1tilde}
\end{align}
We then have:
\begin{align}
\frac{\partial \Delta F_1^0}{\partial n_0}
=\int d^3r \left\{\frac{\delta \Delta \tilde{F}_1^0}{\delta n(r)}\frac{\partial n(r)}{\partial n_0}\right\}
+\frac{\partial \Delta \tilde{F}_1^0}{\partial n_0}
\end{align}
From Eq.~\eqref{eq_dF1_dF1tilde} and the expression of $n(r)$, we have:
\begin{align}
\frac{\delta \Delta \tilde{F}_1^0}{\delta n(r)}=\frac{\delta \Delta F_1^{0}}{\delta q(r)}=\mu_0-\bar{v}(r)
\end{align}
and from Eq.~\eqref{free_energy_difference_integral}, we have:
\begin{align}
\frac{\partial \Delta \tilde{F}_1^0}{\partial n_0}=\int d^3r \left\{ -\frac{\partial f_0^{0}(n_0;T)}{\partial n_0} \right\}  = -\int d^3r \left\{\mu_0 \right\}
\label{df00dn0}
\end{align}  
We thus obtain:	
\begin{align}
&\frac{\partial \Delta F_1^0}{\partial n_0} = - \int d^3r \left\{\bar{v}(r)\right\}
\label{d Delta F_10 / d n_0}	
\end{align}

In the Thomas-Fermi case the proves of Eqs. \eqref{eq_dF10_dq} and \eqref{eq_dF10dn0} are much simpler and will not be given here. 
	
	\section{Useful relations for the virial theorem derivation\label{app_vir_vel_vxc}}
	\subsection{Preliminary calculation}
	Let $f(r)$ and $g(r)$ be two symmetric functions of $r$, whose respective Fourier transforms are $f_k$, $g_k$, have the following properties:
	\begin{align}
		&\lim_{k\rightarrow 0} k f_k=\lim_{k\rightarrow 0} k g_k =0\\
		&\lim_{k\rightarrow \infty} f_k =\lim_{k\rightarrow \infty} g_k =0
	\end{align}
	Consider the following integral:
	\begin{align}
		I\equiv&\int d^3r\left\{f(r) \vec{r}.\nabla_\vec{r} g(r)\right\}
	\end{align}
	Switching to the Fourier space and integrating by part, we may write:
	\begin{align}
		I=&\int d^3r\int \frac{d^3k}{(2\pi)^3}\int \frac{d^3k'}{(2\pi)^3}
		\left\{f_{k'}(-i\vec{k}.\vec{r})g_k e^{-i(\vec{k}+\vec{k}').\vec{r}}\right\}
		\\
		=&-3\int \frac{d^3k}{(2\pi)^3}\left\{f_k g_k\right\}
		-\int \frac{d^3k}{(2\pi)^3}\left\{f_k \vec{k}.\nabla_\vec{k}g_k\right\}
		\label{eq_vir_useful_relation1}
	\end{align}
	Making one more step of integration by part, we get
	\begin{align}
		I=&-3\int \frac{d^3k}{(2\pi)^3}\left\{f_k g_k\right\}
		-\left[\frac{4\pi k^3}{(2\pi)^3} f_k g_k\right]_0^\infty
		\nonumber\\
		&+\int_0^\infty \frac{dk}{(2\pi)^3}\left\{g_k\left(3\cdot 4\pi k^2 f_k+4\pi k^3 \frac{\partial f_k}{\partial k}\right)\right\}
		\\
		=&\int \frac{d^3k}{(2\pi)^3}\left\{g_k \vec{k}.\nabla_\vec{k}f_k\right\}\label{eq_vir_useful_relation2}
	\end{align}
	Both Eqs~\eqref{eq_vir_useful_relation1} and \eqref{eq_vir_useful_relation2} are useful in the virial theorem derivation.

	\subsection{Virial of $v_\text{xc}$\label{app_vir_vxc}}
	
	\begin{align}
		I_\text{xc}^\text{virial}\equiv&\int d^3r\left\{\left(n_0+q(r)\right)\vec{r}.\nabla_\vec{r}v_\text{xc}(n_0+q(r))\right\}
	\end{align}
	Integrating by parts, we have:
	\begin{align}
&I_\text{xc}^\text{virial}=
-3 \int d^3r\left\{
(n_0+q(r))\left(v_\text{xc}(n_0+q(r))-v_\text{xc}(n_0)\right)\right\}
\nonumber\\
&-\int d^3r \left\{ \left( v_\text{xc}(n_0+q(r)) - v_\text{xc}(n_0)\right) r\frac{d}{dr}(n_0+q(r)) \right\}
\nonumber\\
&=-3 \int d^3r\left\{ (q(r) + n_0) -n_0)(v_\text{xc}(n_0) \right\}\nonumber\\
&\hphantom{=}-3 \int d^3r\left\{ (q(r)+n_0)(v_\text{xc}(q(r)+n_0)-n_0v_\text{xc}(n_0) \right\}
\nonumber\\
&\hphantom{=}- \int d^3r\left\{ (v_\text{xc}(q(r) + n_0) - v_\text{xc}(n_0))r\dfrac{d}{dr}(q(r) + n_0) \right\}
\nonumber\\
&\hphantom{=}+ v_\text{xc}(n_0)\int d^3r\left\{ r\dfrac {d}{dr}(q(r) + n_0) \right\}
\nonumber\\
&\hphantom{=}+ \int d^3r\left\{r\dfrac {d}{dr}(f_\text{xc}(q(r) + n_0)-(f_\text{xc}(n_0)\right\}
	\end{align}
The last two terms can be integrated by parts:
\begin{align}
v_\text{xc}(n_0)&\int d^3r\left\{r\frac {d}{dr}\big(q(r) + n_0\big)\right\}
\nonumber\\
&= - 3v_\text{xc}(n_0)\int d^3r\left\{ q(r) \right\}	
\end{align} 
and
\begin{align}
&\int d^3r\left\{ r\frac {d}{dr}\big(f_\text{xc}(q(r) + n_0)-f_\text{xc}(n_0)\big) \right\}\nonumber\\
&= - 3\int d^3r\left\{f_\text{xc}(q(r) + n_0)-f_\text{xc}(n_0)\right\}	
\end{align}
leading to the result :
\begin{align}
&I_\text{xc}^\text{virial}=3\Delta F_1^\text{xc}
\nonumber\\
&-3
\int d^3r\left\{(n_0+q(r))v_\text{xc}(n_0+q(r))-n_0v_\text{xc}(n_0)\right\}
\end{align}

	\subsection{Virial of $v_\text{el}$\label{app_vir_vel}}
	Using Eq.~\eqref{eq_vir_useful_relation2}, we can write:
	\begin{align}
		I_\text{el}^\text{virial}
		\equiv&\int d^3r\left\{\left(n_0+q(r)\right)\vec{r}.\nabla_\vec{r}\tilde{v}_\text{el}(r)\right\}\\
		=&\int \frac{d^3k}{(2\pi)^3}\left\{\tilde{v}_{\text{el},k}\vec{k}.\nabla_\vec{k}q_k\right\}
		+n_0 \int d^3r\left\{\vec{r}.\nabla_\vec{r}\tilde{v}_\text{el}(r)\right\}\label{eq_vir_el_interm0}
	\end{align}
	Integrating by part, the second term is immediately shown to be zero:
	\begin{align}
		n_0 \int d^3r\left\{\vec{r}.\nabla_\vec{r}\tilde{v}_\text{el}(r)\right\}
		=-3n_0 \int d^3r\left\{\tilde{v}_\text{el}(r)\right\}=0
	\end{align}
	Now let us focus on the first term of Eq.~\eqref{eq_vir_el_interm0}. Using the expression of $v_\text{el}(r)$, Eq.~\eqref{eq_def_v_el}, we have:
	\begin{align}
		\int& \frac{d^3k}{(2\pi)^3}\left\{\tilde{v}_{\text{el},k}\vec{k}.\nabla_\vec{k}q_k\right\}
		=\int \frac{d^3k}{(2\pi)^3}\left\{v_{\text{el},k}\vec{k}.\nabla_\vec{k}q_k\right\}
		\nonumber\\
		=&\int_0^\infty \frac{dk}{(2\pi)^3}\left\{(4\pi)^2e^2(1+n_\text{$\nu$}h_k)(Z-q_k)k \frac{\partial q_k}{\partial k}\right\}
		\\
		=&\int_0^\infty \frac{dk}{(2\pi)^3}\left\{(4\pi)^2e^2(Z-q_k)k \frac{\partial q_k}{\partial k}\right\}
		\nonumber\\
		&+n_\text{$\nu$}\int_0^\infty \frac{dk}{(2\pi)^3}\left\{(4\pi)^2e^2 h_k(Z-q_k)k \frac{\partial q_k}{\partial k}\right\}\\
		\equiv& I_\text{el,1}^\text{virial}
		+I_\text{el,2}^\text{virial}\label{eq_vir_el_interm1}
	\end{align}
	For the first term of the latter equation, we integrate by part and get:
	\begin{align}
		I_\text{el,1}^\text{virial}=&\int_0^\infty \frac{dk}{(2\pi)^3}\left\{(4\pi)^2e^2k \frac{\partial }{\partial k}\left(\left(Z-\frac{q_k}{2}\right)q_k\right) \right\}
		\\
		=&\left[ \frac{(4\pi)^2e^2}{(2\pi)^3}k\left(Z-\frac{q_k}{2}\right)q_k\right]_0^\infty
		\nonumber\\
		&-\int_0^\infty \frac{dk}{(2\pi)^3}\left\{(4\pi)^2e^2\left(Z-\frac{q_k}{2}\right)q_k \right\}
		\\
		=&-\int\frac{d^3k}{(2\pi)^3}\left\{\frac{4\pi e^2}{k^2}\left(Z-\frac{q_k}{2}\right)q_k \right\}=-W_\text{intra}
	\end{align}
	
	For the second term of Eq.~\eqref{eq_vir_el_interm1}, we integrate by part and get:
	\begin{align}
I_\text{el,2}^\text{virial}=&
\int_0^\infty \frac{dk}{(2\pi)^3}\left\{(4\pi)^2e^2 n_\text{$\nu$} h_k k \frac{\partial }{\partial k}\left(\left(Z-\frac{q_k}{2}\right)q_k\right) \right\}
\\
=&\left[ \frac{(4\pi)^2e^2}{(2\pi)^3}n_\text{$\nu$} h_k k\left(Z-\frac{q_k}{2}\right)q_k\right]_0^\infty
\nonumber\\
&-n_\text{$\nu$}\int_0^\infty \frac{dk}{(2\pi)^3}\left\{(4\pi)^2e^2\left(Z-\frac{q_k}{2}\right)q_k 
\frac{\partial }{\partial k}\left(k h_k\right)  \right\}
\\
=&-\frac{n_\text{$\nu$}}{2}\int_0^\infty \frac{dk}{(2\pi)^3}\left\{
\vphantom{\frac{\partial }{\partial k}}(4\pi)^2e^2\left(Z^2+2Zq_k -q_k^2\right)
\right.
\nonumber\\
&\hphantom{-\frac{n_\text{$\nu$}}{2}\int_0^\infty \frac{dk}{(2\pi)^3}}\left.\frac{\partial }{\partial k}\left(k h_k\right)  \right\}\\
=&-\frac{n_\text{$\nu$}}{2}\int_0^\infty \frac{dk}{(2\pi)^3}
\left\{4\pi k^2 v_{\text{ii},k} h_k\right\}
\nonumber\\
&-\frac{n_\text{$\nu$}}{2}\int_0^\infty \frac{dk}{(2\pi)^3}
\left\{4\pi k^2 v_{\text{ii},k} k \frac{\partial h_k}{\partial k}\right\}
\end{align}
Using Eq.~\eqref{eq_vir_useful_relation2}, we get:
\begin{align}
I_\text{el,2}^\text{virial}=&-W^\text{i\ approx}-\frac{n_\text{$\nu$}}{2}
\int d^3r\left\{h(r) \vec{r}.\nabla_{\vec{r}}v_\text{ii}(r)\right\}
\end{align}
	
Finally, we have:
\begin{align}
I_\text{el}^\text{virial}
=&-W_\text{intra}
-\frac{n_\text{$\nu$}}{2}\int d^3r\left\{h(r)\vec{r}.\nabla_\vec{r}v_\text{ii}(r)\right\}
\nonumber\\
&-W^\text{i\,approx}
\end{align}
	
	\subsection{Virial theorem for the HNC and DH model of classical fluids\label{app_virial_classical_fluid}}
	The expressions for the fluid free-energy in the HNC and DH models are given in Eqs. \eqref{eq_hnc_free_energy_renorm} and \eqref{eq_dh_free_energy_renorm}, respectively. For the sake of keeping the paper self-contained, we recall below the expression of the excess pressure for these two models:
	\begin{align}
		P^\text{HNC}\equiv& n_\text{$\nu$}^2
		\left.
		\frac{\partial \bar{A}^\text{HNC}}{\partial n_\text{$\nu$}}
		\right|_\text{eq}
		\\
		=&\frac{n_\text{$\nu$}^2}{2\beta}
		\int d^3r\left\{\vphantom{\frac{(R)^2}{2}}
		h(r)\beta v(R)
		\right.\nonumber\\&\left.\left.
		+(h(r)+1)\log\left(h(r)+1\right)
		-h(r)-\frac{h(r)^2}{2}\right\}\right|_\text{eq}
		\nonumber\\&
		+\frac{1}{2\beta}
		\left.\int \frac{d^3k}{(2\pi)^3}\left\{
		\log\left(1+n_\text{$\nu$} h_k\right) - n_\text{$\nu$} c_k
		\right\}\right|_\text{eq}
		\label{eq_press_hnc_model_class_fluid}
	\end{align}
where we used the Ornstein-Zernike (OZ) relation Eq.~\eqref{eq_fluid_integral_oz} in the Fourier space, $c_k$ being the Fourier transform of the direct correlation function $c(r)$.
	\begin{align}
		P^\text{DH}\equiv& n_\text{$\nu$}^2
		\left.
		\frac{\partial \bar{A}^\text{DH}}{\partial n_\text{$\nu$}}
		\right|_\text{eq}
		\\
		=&\frac{n_\text{$\nu$}^2}{2\beta}
		\left.\int d^3r\left\{h(r)\beta v(r)\right\}\right|_\text{eq}
		\nonumber\\&
		+\frac{1}{2\beta}
		\left.\int \frac{d^3k}{(2\pi)^3}\left\{
		\log\left(1+n_\text{$\nu$} h_k\right) +\beta n_\text{$\nu$} v_k
		\right\}\right|_\text{eq}
		\label{eq_press_dh_model_class_fluid}
	\end{align}
	where we use the DH equation Eq.~\eqref{eq_fluid_integral_dh} in the Fourier space, $u_k$ being the Fourier transform of the interaction potential $v(r)$.
	Here, the $|_\text{eq}$ symbol in Eqs~\eqref{eq_press_hnc_model_class_fluid} and \eqref{eq_press_dh_model_class_fluid} means that $h(r)$ fulfills the integral equation corresponding to the chosen model, for the given interaction potential $v(r)$. In the context of Eq.~\eqref{eq_pressure_formula}, $v(r)$ (respectively $v_k$) is indeed $v_{\text{ii}}(r)$ (respectively $v_{\text{ii},k}$). In the following, we will omit the $|_\text{eq}$ symbol to shorten the notation.
	
Let us first consider the following integral:
\begin{align}
I_1^\text{HNC}=
\frac{n_\text{$\nu$}^2}{2\beta}
\int d^3r&\left\{\vphantom{\frac{(R)^2}{2}}
h(r)\beta v(r)
+(h(r)+1)\log\left(h(r)+1\right)
\right.\nonumber\\&\left.
-h(r)-\frac{h(r)^2}{2}\right\}
\end{align}
which corresponds to the first term of Eq.~\eqref{eq_press_hnc_model_class_fluid}. Integrating by part, we get:
\begin{align}
I_1^\text{HNC}=-
\frac{n_\text{$\nu$}^2}{6\beta}
\int d^3r&\left\{\left(\beta v(r)+\log\left(h(r)+1\right)
\right.\right.\nonumber\\&\left.\left.
-h(r)\right)\vec{r}.\nabla_{\vec{r}}h(r)
+h(r)\beta\vec{r}.\nabla_{\vec{r}}v(r)\right\}
\\
=-
\frac{n_\text{$\nu$}^2}{6\beta}
\int d^3r&\left\{-c(r)\vec{r}.\nabla_{\vec{r}}h(r)
+h(r)\beta\vec{r}.\nabla_{\vec{r}}v(r)\right\}
\end{align}
where we used the HNC closure relation Eq.~\eqref{eq_fluid_integral_hnc_closure}. Using Eq.~\eqref{eq_vir_useful_relation1} on the first term, we get:
\begin{align}
I_1^\text{HNC}
=&-\frac{n_\text{$\nu$}^2}{2\beta}
\int\frac{d^3k}{(2\pi)^3}\left\{c_kh_k\right\}
-\frac{n_\text{$\nu$}^2}{6\beta}
\int \frac{d^3k}{(2\pi)^3}\left\{c_k\vec{k}.\nabla_{\vec{k}}h_k\right\}
\nonumber\\
&-\frac{n_\text{$\nu$}^2}{6}
\int d^3r\left\{h(r)\vec{r}.\nabla_{\vec{r}}v(r)\right\}
\label{eq_virial_I1HNC_term}
\end{align}

Let us now consider the following integral:
\begin{align}
I_1^\text{DH}=
\frac{n_\text{$\nu$}^2}{2\beta}
\int d^3r&\left\{h(r)\beta v(r)\right\}
\end{align}
which corresponds to the first term of Eq.~\eqref{eq_press_dh_model_class_fluid}. Performing the same operations, we obtain:
\begin{align}
I_1^\text{DH}=&
-\frac{n_\text{$\nu$}^2}{6\beta}
\int d^3r\left\{\beta v(r) \vec{r}.\nabla_{\vec{r}}h(r)+h(r)\beta \vec{r}.\nabla_{\vec{r}}v(r)\right\}
\\
=&
\frac{n_\text{$\nu$}^2}{2\beta}
\int \frac{d^3k}{(2\pi)^3}\left\{\beta v_k h_k\right\}
+\frac{n_\text{$\nu$}^2}{6\beta}
\int \frac{d^3k}{(2\pi)^3}\left\{\beta v_k \vec{k}.\nabla_{\vec{k}}h_k\right\}
\nonumber\\
&-\frac{n_\text{$\nu$}^2}{6}
\int d^3r\left\{h(r)\vec{r}.\nabla_{\vec{r}}v(r)\right\}
\label{eq_virial_I1DH_term}
\end{align}

Let us finally consider the integral:
\begin{align}
I_2=\frac{1}{2\beta}
\int \frac{d^3k}{(2\pi)^3}\left\{
\log\left(1+n_\text{$\nu$} h_k\right)
\right\}
\end{align}
which is common to Eqs~\eqref{eq_press_hnc_model_class_fluid} and \eqref{eq_press_dh_model_class_fluid}. Integrating by part, we get:
\begin{align}
I_2=&-\frac{n_\text{$\nu$}}{6\beta}
\int \frac{d^3k}{(2\pi)^3}\left\{
\frac{1}{1+n_\text{$\nu$} h_k}\vec{k}.\nabla_{\vec{k}}h_k
\right\}
\\
=&\begin{cases}
-\frac{n_\text{$\nu$}}{6\beta}
\int \frac{d^3k}{(2\pi)^3}\left\{
(1-n_\text{$\nu$}c_k)\vec{k}.\nabla_{\vec{k}}h_k
\right\}&\text{ (HNC case)}
\\
-\frac{n_\text{$\nu$}}{6\beta}
\int \frac{d^3k}{(2\pi)^3}\left\{
(1+n_\text{$\nu$}\beta v_k)\vec{k}.\nabla_{\vec{k}}h_k
\right\}&\text{ (DH case)}
\end{cases}
\end{align}
where we used again the Ornstein-Zernike (OZ) relation in the HNC case and the DH equation in the DH case. Integrating by part the first term, we obtain:
\begin{align}
I_2
=&-\frac{n_\text{$\nu$}}{2\beta}\int \frac{d^3k}{(2\pi)^3}\left\{h_k\right\}
\nonumber\\
&+\begin{cases}
\frac{n_\text{$\nu$}^2}{6\beta}
\int \frac{d^3k}{(2\pi)^3}\left\{
c_k\vec{k}.\nabla_{\vec{k}}h_k
\right\}&\text{ (HNC case)}
\\
\frac{n_\text{$\nu$}^2}{6\beta}
\int \frac{d^3k}{(2\pi)^3}\left\{
-\beta v_k\vec{k}.\nabla_{\vec{k}}h_k
\right\}&\text{ (DH case)}
\end{cases}
\label{eq_virial_I2_term}
\end{align}

Using Eqs~\eqref{eq_virial_I1HNC_term}, \eqref{eq_virial_I2_term} in \eqref{eq_press_hnc_model_class_fluid} plus the OZ relation for the HNC case, and Eqs~\eqref{eq_virial_I1DH_term},\eqref{eq_virial_I2_term} in \eqref{eq_press_dh_model_class_fluid} plus the DH equation for the DH case, we get:
\begin{align}
P^{\left|\substack{\text{HNC}\\\text{DH}}\right.}=-\frac{n_\text{$\nu$}^2}{6}
\int d^3r\left\{h(r)\vec{r}.\nabla_{\vec{r}}v(r)\right\}
\end{align}
which corresponds to the virial theorem for a simple fluid with arbitrary interaction potential.

	\bibliographystyle{unsrt}
	\bibliography{bibexport.bib}
	
\end{document}